\documentclass[12pt,a4paper]{article}

\usepackage{cite}
\usepackage{cite}

\usepackage{comment}

\usepackage{mathtools}

\usepackage{latexsym,amsfonts,amsmath,amssymb}
\usepackage{bbm}
\usepackage{hyperref}
\usepackage{empheq}
\usepackage{graphicx}
\usepackage{color}
\usepackage{caption}
\usepackage{subcaption}
\usepackage[normalem]{ulem}
\usepackage{comment}
\usepackage{url}
\usepackage{slashed}
\usepackage{tabu}
\usepackage{multirow}
\usepackage{pifont}

\usepackage{mathtools}

\usepackage{amsxtra,graphics,epsfig,bm,tikz,xfrac,lscape}
\usetikzlibrary{decorations.pathmorphing}
\usetikzlibrary{decorations.markings}
\usetikzlibrary{arrows, decorations.markings, calc, fadings, decorations.pathreplacing, patterns, decorations.pathmorphing, positioning}

\usetikzlibrary{positioning,shapes}
\usetikzlibrary{chains}
\usetikzlibrary{arrows,fit,decorations.pathreplacing}
\tikzstyle{every picture}+=[remember picture]
\tikzstyle{na} = [baseline=-.5ex]

\makeatletter \@addtoreset{equation}{section} \@addtoreset{equation}{subsection} \makeatother

\newcommand\cI{{\cal I}}

\newcommand\cL{{\cal L}}

\newcommand\cN{{\cal N}}

\def\ads2{AdS$_2$}

\newcommand{\Dslash}{\; \ensuremath \raisebox{0.03cm}{\slash}\hspace{-0.28cm} D}

\newcommand{\nn}{\nonumber}

\def\i{\mathrm{i}}

\def\beqa{\begin{eqnarray}\displaystyle}
\def\eeqa{\end{eqnarray}}
\def\be{\begin{equation}}
\def\ee{\end{equation}}
\newcommand{\ba}{\begin{array}}
\newcommand{\ea}{\end{array}}

\def\bse{\begin{subequations}}
\def\ese{\end{subequations}}

\newcommand{\bem}{\begin{pmatrix}}
\newcommand{\eem}{\end{pmatrix}}

\renewcommand{\=}{\;  = \;}
\def\+{\, + \,}

\def\wt{\widetilde}

\def\bar{\overline}
\def\ads2{AdS$_2$}
\def\ss2{S$^2$}

\def\rt2{\sqrt{2}}

\newcommand\qeq{Q_{eq}}

\def\g{\gamma}

\def\a{\alpha}
\def\b{\beta}

\def\k{\kappa}
\def\eps{\epsilon}

\def\l{{\lambda}}

\def\ve{\varepsilon}

\def\r{ r}

\def\S{{\sigma}}
\def\P{{\rho}}

\def\scvrho{{\rho}}

\def\l{\lambda}

\def\i{\mathrm{i}}
\def\a{\alpha}

\def\sc{\phi}
\def\ads2{AdS$_2$}
\def\eads2{EAdS$_2$}

\def\i{\mathrm{i}}

\def\ZZZ{{\hbox{ Z\kern-1.6mm Z}}}
\def\RRR{{\hbox{ R\kern-2.4mm R}}}
\def\CCC{{\hbox{ C\kern-2.0mm C}}}
\def\zzz{{\hbox{z\kern-1mm z}}}

\def\one{{\hbox{ 1\kern-.8mm l}}}
\def\zero{{\hbox{ 0\kern-1.5mm 0}}}

\newcommand{\bea}[1]{\begin{eqnarray}\label{#1} }
\newcommand{\eea}{\end{eqnarray}}

\begin{document}
\numberwithin{equation}{subsection}
\baselineskip 24pt

\begin{center}

{\Large \bf  Supersymmetric spectrum for vector multiplet  \\
on Euclidean AdS$_2$ }

\end{center}

\vskip .6cm
\medskip

\vspace*{4.0ex}

\baselineskip=18pt

\centerline{\large \rm Alfredo Gonz\' alez Lezcano$^{a}$, Imtak Jeon$^{a,b}$  and Augniva Ray$^a$}

\vspace*{4.0ex}

\centerline{\large \it  $^a$ Asia Pacific Center for Theoretical Physics, Postech, Pohang 37673, Korea }

\centerline{\large \it  $^b$ Department of Physics, Postech, Pohang 37673, Korea}

\vspace*{1.0ex}
\centerline{\small E-mail: alfredo.gonzalez@apctp.org,~  imtakjeon@gmail.com, ~augniva.ray@apctp.org}

\vspace*{5.0ex}

\centerline{\bf Abstract} \bigskip
Quantum study of supersymmetric theories on Euclidean two dimensional anti-de Sitter space (\eads2) requires complexified spectrum. For a chiral multiplet, we showed that the spectrum of the Dirac operator acquires a universal shift of~$\i/2$ from the real spectrum to make the supersymmetry between boson and fermion manifest, where both the bosonic and fermionic eigenfunctions are normalizable using an appropriate definition of Euclidean inner product. 
We extend this analysis to the vector multiplet, where we show that
the gaugino requires both $+\i/2$ and $-\i/2$ shift from the real spectrum, and there is additional isolated point at vanishing spectral parameter which is mapped by supersymmetry to the boundary zero modes of the vector field. Furthermore, this spectral analysis shows that not every bosonic fields in the vector multiplet can satisfy normalizable boundary condition.
Nevertheless, aided by a reorganization of fields into a cohomological form, we find the supersymmetry mapping between bosons and fermions in terms of the expansion coefficients with respect to the newly constructed basis. 

\vfill \eject

\tableofcontents

\label{sec:introduction}
The study of quantum aspects of supersymmetric theories on \eads2 \footnote{
Our interest is in two dimensional hyperbolic space $\mathbb{H}_2$ having disc topology as it is relevant geometry for the study of black hole entropy \cite{Sen:2008vm}. Throughout this paper,  by \eads2  we mean the $\mathbb{H}_2$. For study of asymptotic boundary condition on global AdS$_2$ we refer to \cite{correa:2019rdk}. } has received a lot of attention in the last few decades. Most of all because the universal presence of an \ads2 factor in the near horizon geometry of all extremal black holes makes the study of such a geometry an important endeavor to better understand black hole physics. In particular, their thermodynamic properties are studied via the Euclidean path integral approach, where Sen's quantum entropy formalism \cite{Sen:2008vm} has provided us with a powerful insight into quantum properties of extremal black holes \cite{Gibbons:1976ue, Sen:2008vm, Sen:2011ba, Bhattacharyya:2012wz,Sen:2012cj, Sen:2012dw, Keeler:2014bra,Karan:2020njm, Banerjee:2021pdy,Assel:2016pgi,GonzalezLezcano:2023cuh}. 
Furthermore, in the presence of supersymmetry, it is possible to exploit the powerful tool of supersymmetric localization to capture all the quantum corrections achieving exact match with microstate counting of black holes \cite{Banerjee:2009af, Dabholkar:2010uh, Dabholkar:2011ec, Gupta:2012cy, Dabholkar:2014wpa, LopesCardoso:2022hvc, Hristov:2021zai, deWit:2018dix,Jeon:2018kec,Iliesiu:2022kny,Gupta:2021roy,Ciceri:2023mjl}. Study of localization for field theory on AdS background can also be found in \cite{David:2018pex, David:2019ocd,Pittelli:2018rpl,Iannotti:2023jji,GonzalezLezcano:2023cuh,Cabo-Bizet:2017jsl}.

However, this impressive success has suffered from an underlying conceptual problem regarding the preservation of supersymmetry itself in hyperbolic spaces. As noticed in \cite{David:2018pex, David:2019ocd} and highlighted in \cite{Sen:2023dps}, the bosonic and fermionic quantum fluctuations are not mapped to each other by supersymmetry when they are expressed in terms of the standard delta-function  normalizable eigenfunctions 
\cite{Camporesi:1994ga,Camporesi:1992tm,Camporesi:1995fb}  of their respective kinetic operators, due to the asymptotic growth of the Killing spinors. This problem can be sketched by looking at the generic supersymmetry relation between a bosonic field $\Phi$ and a fermionic field $\Psi$, which are related by a local transformation through the Killing spinor $\varepsilon$ as 
\be\ba{l}
Q \Phi = {{\ve}} \Psi\,. \nonumber
\ea\ee
Suppose we consider a scalar and a Dirac spinor as the boson and fermion respectively. When using the standard delta-function normalizable basis functions, both boson and fermion behave as $\exp(-\eta/2)$ for large $\eta$, whereas the Killing spinors behave as $\exp(\eta/2)$ (see the equations  \eqref{AdS2eigenfunction}, \eqref{EigenAdS2a} and \eqref{eq:KillingSpinors}). If we refer to the coefficient of $\eta$ in the exponent of the asymptotic behavior as ``degree of growth'', then we see that the left and right-hand side of the supersymmetry relation have different degree of growth, which are $-1/2$ and $0$ respectively. This mismatch brings into question 
 the very meaning of supersymmetric theory on \eads2. If we consider only the normalizable boundary condition for boson and fermion for a well-defined path integral, then they would not respect supersymmetry. Therefore, the successful 1-loop test of the quantum  entropy with respect to its supersymmetric microscopic counterpart and further exact test using the supersymmetric localization seems to lose its ground.

In~\cite{GonzalezLezcano:2023uar}, this problem was elaborated on  and a resolution was provided for the supersymmetry relation between scalar and spinor fields on \eads2.
It was revealed that a complexified spectrum is required to recover supersymmetry. Upon shifting the spectral parameter of the Dirac operator from real $\lambda$ to $\lambda +\i/2$, the corresponding eigenfunctions of spinor fields manifestly maps to the scalar eigenfunctions. Dealing with the complexified spectrum requires an appropriate definition of the inner product, 
\beqa
\langle \psi^\pm_{\lambda + \frac{\i}{2}\, k} | \psi^\pm_{\lambda' + \frac{\i}{2}\, k'} \rangle 
& \equiv & \pm \i \int {\rm d}\eta {\rm d}\theta \sqrt{g}\, \bigl(\psi^\mp_{\lambda + \frac{\i}{2},k}\,\bigr)^T C\, \psi^\pm_{\lambda' +\frac{\i}{2},k'} 
= \delta(\lambda - \lambda')\delta_{k,k'}\,,\nn
\eeqa
where we define the dual of a basis element $\psi^{\pm}_{\l, k}$ for the inner product through the symplectic Majorana conjugate as $\pm \i (\psi^{\mp}_{\l,k})^T C$ with the charge conjugation matrix $C$ and not through the hermitian conjugate  (see the definition of this basis elements in \eqref{EigenAdS2a}). This is in fact the natural definition of inner product because in Euclidean space the conjugate of a spinor is treated as an independent spinor not being related by complex conjugation~\cite{Osterwalder:1973zr}, and further it can be read off from the kinetic term of the Euclidean action. Using this definition, the fermionic eigenfunctions are delta-function normalizable, resolving the issue of incompatibility between normalizable basis and supersymmetry. 

Our goal now is to extend this discussion to include vector fields, for which we focus on the vector multiplet.
Using a similar procedure of modifying the spectral parameter of gaugini, we construct the supersymmetric basis for quantum fluctuations of a vector multiplet on \eads2 to ensure the supersymmetric mapping between bosons and fermions within the vector multiplet.
The present work goes beyond \cite{GonzalezLezcano:2023uar} in some significant ways:
     \begin{itemize}
         \item In contrast with the case of the chiral multiplet, where the shifting of the spectral parameter was universally $+\frac{\i}{2}$ for all fermionic modes, the case of the vector multiplet requires the spectral parameter to be shifted in two opposite directions in the imaginary axis: $\pm \frac{\i}{2}$ for $\lambda \neq 0$.
         \item   Furthermore, there is an isolated point in the spectrum of the gaugini at $\l =0$, which is left untouched under the complex shift. This allows us to build on \cite{GonzalezLezcano:2023cuh} to identify the superpartner of the boundary zero modes of vector field.
           \item The implication of this spectral analysis is that unlike the chiral multiplet, not every bosonic field in vector multiplet can have normalizable boundary condition.
     \end{itemize}
  The first two points have been depicted in figure \ref{fig:theonlyfigure} and discussed in the subsequent sections. 
\\

The rest of the paper is organized as follows. In section \ref{sec:Problemsofsupersymmetry}, 
we elaborate the problem of supersymmetry relation with the standard basis functions by revisiting the discussion on supersymmetry transformation of the scalar and extending the discussion to the supersymmetry of the vector field. 
 We show how a complexified spectrum of fermionic field cures the problem of supersymmetry, where the structure of the spectrum of gaugino is compared to the spectrum of spinor field of the superpartner of scalar.  
In section \ref{sec:Resolutionoftheproblems}, we present the supersymmetric spectrum of the fields in the chiral multiplet as well as in the vector multiplet and we show the explicit map between mode expansion coefficients of the fields related by supersymmetry. In section ~\ref{sec:Sumaryanddiscussions}, we present a discussion on what we have achieved in the case of the vector multiplet, as well as highlight the present and future directions that might be pursued following our resolution of the problem. Technical digressions as well as details of computations backing our claims are relegated to the Appendices \ref{appendixA} and~\ref{appendixB}.       
\section{Problems of supersymmetry on  \eads2} \label{sec:Problemsofsupersymmetry}
In this section we state the problem of supersymmetry on \eads2  associated with the standard delta-function normalizable eigenfunctions. We revisit this discussion for the case of supersymmetry of scalar field, labelled by $\Phi$, that was analyzed in~\cite{GonzalezLezcano:2023uar}, and extend it to  the case of  supersymmetry of the vector field $A_\mu$. This analysis provides the resolution $via$ complexification of the fermionic spectrum.
To this end, we begin by reviewing the eigenfunctions for scalar, vector and spinor fields~\cite{Camporesi:1994ga, Camporesi:1992tm, Camporesi:1995fb
} as well as their main properties that will be useful in our discussion.

\subsection{Eigenfunctions}
On the \eads2 background of unit radius whose metric is given by
 \begin{align} \label{adsmetric}
 ds^2  =  d\eta^2 + \sinh^2\eta\, d\theta^2\,,  
 \end{align} 
 with the range $0\leq \eta < \infty\,, ~~ 0\leq \theta <2\pi\,$, the basis functions for the scalar, vector and spinor fields are given as follows. 
 
\paragraph{Scalar modes:} 
For the Laplacian operator  $-\nabla^2$ on \eads2, the complete set of  eigenfunctions is given by $\{ \phi_{\l,k} \,|\,\l \in \mathbb{R}_{>0}\,, k\in \mathbb{Z}\}$, whose explicit form  is 
\begin{align}
\begin{split}
\label{AdS2eigenfunction}
\phi_{\lambda, k}(\eta,\theta) & =  A_{\l,k} {\rm e}^{\i k \theta} 
 {\sinh}^{| k|}\eta\,  F\left(\frac{1}{2}+| k|+\i\lambda, \frac{1}{2} +| k|  - \i \lambda
 ; | k| +\!1; - \!\sinh^2\!\frac{\eta}{2} \right)\, , 
 \end{split}
\end{align}
where $F$ denotes the hypergeometric function and $A_{\l,k}$ is a normalization constant
\begin{align}
\begin{split}
 A_{\l,k} & =  \frac{1}{\sqrt{2\pi }}\frac{1}{2^{| k|}| k|!} \!\left(\! \frac{\Gamma(\frac{1}{2}+| k| +\i\lambda)\Gamma(\frac{1}{2}+| k| -  \i \lambda )}{\Gamma(\i\lambda)\Gamma(-\i\lambda)} \right)^{\!\frac{1}{2}} \,.\nn 
\end{split}
\end{align}
The eigenfunctions have the following symmetry $\phi_{-\l,k} = \phi_{\l,k}$\,, and their eigenvalues are given by
\begin{align} \label{eq:klambda}
    - \nabla^2 \phi_{\lambda, k} & =
    \left(\l^2 + \frac{1}{4}\right)  ~\phi_{\lambda, k}\,\, . 
\end{align}
Furthermore, the eigenfunctions~\eqref{AdS2eigenfunction} satisfy delta-function orthonormality under the following definition of inner product
\begin{align}\label{scalarinnerprod}
\big\langle \sc_{\lambda,k} \big| \sc_{\lambda',k'}\big\rangle  \equiv \int  {\rm d}\eta {\rm d}\theta \sqrt{g}\,\sc_{\lambda,-k}
\sc_{\lambda',k'}
= 
\delta(\lambda -\lambda')\delta_{k,k'}\,.
\end{align}
Here,  we have defined the dual of a basis element $\sc_{\l, k}$ for the inner product as $\sc_{\l, -k}$, stemming from the property of the basis function that 
$\sc_{\lambda , - k}  = (\sc_{\lambda ,  k})^\ast \,
$
for real $\lambda$. Following the spirit of the Euclidean theory, we deliberately avoid using the complex conjugation for the inner product. 

The spectral density can be obtained using homogeneity of \eads2 by evaluating the eigenfunctions  at $\eta=0$, where only the $k=0$ mode survives. Hence 
\be\ba{l} \label{eq:muscalar}
\mu_{\sc}(\lambda)\equiv \sum_k \bigl( \sc_{\lambda,-k}\,\sc_{\lambda,k}\bigr)\,\bigr|_{\eta=0}\= \frac{\lambda}{2\pi}\tanh \pi \lambda ~\delta_{k,0}\,.
\ea\ee
We note that $\l =0$ is not part of the scalar spectrum since $\mu_\phi (0) =0$.
By using the inversion formula of hypergeometric function~\eqref{Inverse2F1}, we can show that  
the eigenfunctions asymptotically behave as $\eta \rightarrow \infty$ in the following way
\be \label{scalarasymp}\ba{l}
\sc_{\lambda ,  k}(\eta, \theta)
\;\sim\;  {\rm e}^{-\frac{1}{2}\eta + \i k\theta}  \left( \wt{A}_{\lambda,k} {\rm e}^{\i \lambda\eta }+ \wt{A}_{-\lambda,k} {\rm e}^{-\i \lambda \eta } \right)
\,,
\ea\ee
where $\widetilde{A}_{\l, \,k}$ is a function of $\l$ and $k$,
\be
\wt{A}_{\lambda,k}= \frac{1}{\pi \sqrt{2}}\left( \frac{\Gamma(\i\lambda)\Gamma(\frac{1}{2} +|k| -\i\lambda )}{\Gamma(-\i\lambda)\Gamma(\frac{1}{2} +|k| +\i\lambda )} \right)^{\!\frac{1}{2}}\,.\nn
\ee
\paragraph{Vector modes:} For the Laplacian operator  $-\nabla^2$ acting on vectors defined on \eads2, the eigenfunctions consist of a mixed discrete and continuous spectrum $\{U^{(i)}_{\mu\,\l,k}\,
|\,i=1,2\,; \l \in \mathbb{R}_{>0}; k\in \mathbb{Z} \}$
together with a purely discrete spectrum $\{ \partial_\mu \Lambda^{(\ell)} \,|\,\ell = \pm 1, \pm2, \cdots \}$. The eigenfunctions with the continuous spectrum can be constructed from the scalar eigenfunctions $\phi_{\l,k}$ 
given in \eqref{AdS2eigenfunction} as:
\begin{equation}\label{eq:AdS2VecEigenfunction}
    U^{(1)}_{\mu,\l,k} :=\frac{1}{\sqrt{\k_\l}} \partial_{\mu} \phi_{\lambda, \, k}\,,\qquad U^{(2)}_{\mu,\l,k}:= \frac{1}{\sqrt{\k_\l}} \epsilon_\mu^{~ \nu} \partial_{\nu} \phi_{\lambda, \, k}\,, 
 \end{equation}  
with $\k_\l = \l^2 + \frac{1}{4}$. As the form suggests, they represent respectively the degrees of pure gauge modes and the dynamical mode.   They have eigenvalues as 
\be
    - \nabla^2 U^{(i)}_{\mu,\l,k}  = 
    \left(\l^2 +\frac{5}{4}\right) U^{(i)}_{\mu,\l,k} \,,
\ee
and satisfy the delta-function orthonormality condition
\begin{align}
 \big\langle U^{(i)}_{\mu,\lambda,k} \Big| U^{(j)\mu}{}_{\lambda',k'}\big\rangle    &\; \equiv\; \int \! {\rm d}\eta {\rm d}\theta \sqrt{g}\,U^{(i)}_{\mu,\lambda,-k} U^{(j)\mu}{}_{\lambda',k'}
\= 
\delta^{ij}\delta(\lambda \!-\!\lambda')\delta_{k,k'}\,.
\end{align}

The spectral density of the vector modes can be obtained using the eigenfunctions evaluated at $\eta=0$, where only the $k=1$ mode survives. Explicitly, 
\begin{eqnarray}
&&\mu^{(i)}_A (\l)  =\sum_{k } U^{(i)}_{\mu, \l,k}~U^{(i)\mu}{}_{\l,-k}\Big|_{\eta\rightarrow 0} = \frac{\lambda  \sinh (\pi  \lambda ) \Gamma \left(\frac{3}{2}-\i \lambda \right) \Gamma \left(\frac{3}{2} + \i \lambda\right)}{4 \pi ^2  \left(\l^2+\frac{1}{4}\right)}  \, .  
\end{eqnarray}
Here too $\l =0$ is not part of the spectrum since $\mu^{(i)}_A (0) =0$.

The additional eigenfunctions with discrete spectrum, which we call ``boundary zero modes'',  are given as
\beqa \label{nonnormalizablescalarsads2}
\partial_\mu \Lambda^{(\ell)}\,, \quad \Lambda^{(\ell)} 
\= \dfrac{1}{\sqrt{2\pi |\ell|}} \Biggl(\dfrac{\sinh \eta }{1+\cosh \eta }\Biggr)^{|\ell|} {\rm e}^{i \ell \theta} \, , \quad \ell = \mathbb{Z}- \{0\}  \,,
\eeqa
where, the parameters $\Lambda^{(\ell)}$ are not normalizable as their asymptotic behavior is  $   \Lambda^{(\ell)} \sim O(1)$ for large $ \eta\,$, surviving at the boundary of \eads2.  In fact this parameter is related to the eigenfunction of scalar by $\Lambda^{(\ell)}= - \i \sqrt{2\pi}~\phi_{\pm \frac{\i}{2},\ell}$. These modes cannot be gauged away and their physical contribution should be taken into account \cite{Sen:2011ba}. The asymptotic behavior of the vector eigenfunctions
follows from the asymptotic behavior of the scalar given in \eqref{scalarasymp}. 

As the parameter \eqref{nonnormalizablescalarsads2} satisfies $-\nabla^2 \Lambda^{(\ell)}=0$, the boundary  zero modes have the following eigenvalue equation
\be
-\nabla^2 \partial_\mu \Lambda^{(\ell)}\=   \partial_\mu \Lambda^{(\ell)}\,.
\ee
Moreover they also satisfy the orthogonality condition
\beqa
\big\langle \partial_{\mu}\Lambda^{(\ell)} \Big| \partial^{\mu}\Lambda^{(\ell')}\big\rangle 
 & \equiv &\int \! {\rm d}\eta {\rm d}\theta \sqrt{g}\,
\partial_{\mu}\Lambda^{(-\ell)} \,\partial^{\mu}\Lambda^{(\ell')}
\=\delta_{\ell,\ell'}\,.
\eeqa
and are orthogonal to the other eigenfunctions given in \eqref{eq:AdS2VecEigenfunction}. 

\paragraph{Spinor modes:} 
For the Dirac operator $\i \slashed{D}
$ on \eads2,  
the complete set of eigenfunctions is given by $\{\psi^\pm_{\l,k}\,| \, \l \in \mathbb{R}\,, k \in\mathbb{Z}_{\geq 0} \}$, whose explicit form 
with the gamma matrices being Pauli matrices is  
\be \ba{l}\label{EigenAdS2a} 
\psi^+_{\l, k}\left(\eta,\theta\right)=\! B_{\l, \, k} {\rm e}^{\i(k+\frac{1}{2})\theta}
\begin{pmatrix}   \cosh^{k+1}\frac{\eta}{2}\sinh^{k}\frac{\eta}{2}\, F(k\!+\!1\!+\!\i\lambda, k\!+\!1\!-\!\i\lambda; k\!+\!1;-\sinh^2\!\frac{\eta}{2})\\ 
 -\i\frac{\lambda}{k+1} \cosh^{k}\!\frac{\eta}{2}\sinh^{k+1}\!\frac{\eta}{2}\,F(k\!+\!1\!+\!\i\lambda, k\!+\!1\!-\!\i\lambda; k\!+\!2; -\sinh^2\!\frac{\eta}{2})\end{pmatrix}, 
\\ 
\vspace{0.1mm} \\
\psi^-_{\lambda,k}\left(\eta,\theta\right)=B_{\l, \, k} {\rm e}^{-\i(k+\frac{1}{2})\theta} 
\!\begin{pmatrix}\i \frac{\lambda}{k+1}  \cosh^k\frac{\eta}{2}\sinh^{k+1}\!\frac{\eta}{2}\,F(k\!+\!1\!+\!\i\lambda, k\!+\!1\!-\!\i\lambda; k\!+\!2;-\sinh^2\frac{\eta}{2})\\ 
-\cosh^{k+1}\frac{\eta}{2}\sinh^{k}\!\frac{\eta}{2}\,F(k\!+\!1\!+\!\i\lambda, k\!+\!1\!-\!\i\lambda; k\!+\!1; -\sinh^2\frac{\eta}{2})\end{pmatrix}, 
\ea\ee
where the normalization constant $B_{\l,k}$ is given by
\be
B_{\l,k}=\frac{1}{\sqrt{4\pi }} \frac{1}{k !}\!\left( \frac{\Gamma(1+k+\i\lambda) \Gamma(1+k -\i \lambda)}{\Gamma\left(\frac{1}{2} +\i\lambda\right)\Gamma\left(\frac{1}{2}-\i\lambda\right)}\right)^{\!\frac{1}{2}}\,.\nn
\ee
Note that the ranges of $\l$ and $k$ for spinor field are 
different from those for scalar or vector basis, which are  real numbers and non-negative integers respectively.~\footnote{ We follow the conventions of \cite{GonzalezLezcano:2023uar} which are related to those used in \cite{GonzalezLezcano:2023cuh} as follows: The fermionic basis is organized using spinors $\chi^{\pm}_{\l. k}$ and $\eta^{\pm}_{\l,k}$, we have 
$\psi^+_{\pm|\l| ,k}  = \eta^\mp_{|\l|,k} $ and $
\psi^-_{\pm|\l| ,k} = \pm\chi^{\mp}_{|\l|,k} \,$.} 
The eigenfunctions have the following eigenvalue 
\begin{align}
    \i \Dslash \psi^{\pm}_{\l, k} & =  \l \psi^\pm_{\l, k}\,, 
\end{align}
and they satisfy the delta-function orthonormality condition as
\begin{align}\label{innerprodfermion}
\bigl\langle \psi^\pm_{\lambda,k} \big| \psi^\pm_{\lambda',k'}\big\rangle \equiv \pm \i \int {\rm d}\eta {\rm d}\theta \sqrt{g} \,  \psi^\mp_{\lambda,k} \,\psi^\pm_{\lambda',k^\prime} =\delta(\lambda -\lambda')\delta_{k,k'}\,.
\end{align}
Here, we note again that we define the dual of a basis element $\psi^{\pm}_{\l, k}$  for the inner product by using the symplectic Majorana conjugate $\pm \i (\psi^{\mp}_{\l,k})^T C$\,, where, to avoid clutter, we use the convention that for  spinorial multiplication $\psi \chi \equiv \psi^T C \chi $ with $C =\g_2$.  This definition of dual element stems from the reality condition of the eigenfunction \eqref{EigenAdS2a} $\left(\psi^{\pm}_{\l, k}\right)^\dagger \equiv \pm \i (\psi^{\mp}_{\l,k})^T C$ for real $\l$, but we do not use Hermitian conjugation in the definition of inner product following the spirit of dealing the fermion in Euclidean space \cite{Osterwalder:1972vwp,Osterwalder:1973zr}.  
 Every other combination of inner product vanishes naturally from the $\theta$ integral as $k$ is constrained to be positive. 

Similar to the scalar case, the spectral density can be obtained using homogeneity of \eads2 by evaluating the eigenfunctions at $\eta=0$, where only the $k=0$ mode survives. Hence 
\beqa\label{spectralF}
\mu_{\psi^\pm}(\lambda) \equiv   \pm \i \sum_k  {\psi_{\lambda, k}^\mp} \,\psi_{\lambda,k}^\pm \,\Bigr|_{ \eta=0} =\frac{1}{4\pi} \lambda \coth\pi\lambda \,.
\eeqa
We note that, unlike the case for scalar modes, the spectrum exists at $\lambda=0$ since $\mu_{\psi^{\pm}}(0) \neq 0 $.
Finally, we note that the asymptotic behavior for large $\eta$ is as follows  
\beqa \label{Fermionasympt}
\psi^{\pm}_{\l,k} \!&\sim & {\rm e}^{- \frac{\eta}{2}\pm \i(k+\frac{1}{2})\theta}   \left( {\rm e}^{\i \lambda \eta} \wt{B}_{\lambda,k}  \biggl( \ba{c} 1 \\  -1  \ea \biggr)  \pm {\rm e}^{-\i \lambda \eta} \wt{B}_{-\lambda,k}  \biggl( \ba{c} 1 \\  1  \ea \biggr)\right) ,
\eeqa

where $\wt{B}_{\l, \,k}$ is a function of $\l$ and $k$ whose explicit form is
\be
\wt{B}_{\lambda,k} \=  \frac{1}{2\pi} 
\biggl(\frac{\Gamma\left(\frac{1}{2}+\i \l\right)\Gamma\left(1+k-\i \l\right)}{\Gamma\left(\frac{1}{2}-\i \l\right)\Gamma\left(1+k+\i \l\right)}\biggr)^{\!\frac{1}{2}}  \,.
\ee

\subsection{Supersymmetry of scalar and vector
} \label{sec:issue with Hilbert Space}

The essence of the problem of the standard delta-function normalizable mode presented in previous section with the supersymmetry is due to the asymptotic growth of the Killing spinors. 
The Killing spinors on \eads2 satisfy the following equation,
\begin{align}\label{KSEAdS2}
 D_{\mu} \varepsilon  =  \frac{1}{2} \gamma_{\mu} \varepsilon\,,
\end{align} 
which has four real solutions giving rise to $\cN =(2,2)$ supercharges generating $SL(2,\mathbb{R})$ isometry. For the purpose of our analysis it will be sufficient to focus on $\cN=(1,1)$ subsector associated the following Killing spinors
\be\ba{l} \label{eq:KillingSpinors}
 \epsilon = {\rm e}^{ \tfrac{\i\theta}{2}}\Biggl( \ba{c} \cosh \frac{\eta}{2} \\
 \sinh \frac{\eta}{2} \ea \Biggr), \, \qquad  
 \bar{\epsilon} ={\rm e}^{- \tfrac{\i\theta}{2}}\Biggl(\ba{c}  \sinh \tfrac{\eta}{2}\\
 \cosh \frac{\eta}{2} \ea  \Biggr)\,,
\ea\ee
which generate the Killing vector $v^\mu \,\partial_\mu \equiv \bar\epsilon\gamma^\mu \epsilon\, \partial_\mu = \partial_\theta$ associated to the $U(1)$ subgroup of the $SL(2,\mathbb{R})$ isometry.

We note that these spinors have exponential growth as $\exp({\eta /2})$  for large $\eta$, i.e. they have degree of growth $1/2$.  
We immediately see the mismatch of the asymptotic growth in left- and right-hand sides of the supersymmetry transformation of scalar in \eqref{eq:Qchiral} and of vector in \eqref{Qvector} when expressed in terms of the eigenfunctions of the previous section (except for the supersymmetry of boundary zero modes).  In what follows, we will present the problem in terms of the coefficient in the eigenmode expansion in such a way that not only it makes the problem more explicit but it also suggests an immediate solution.

\subsubsection{Supersymmetry of scalar}\label{SUSYScalar}
Generically, the supersymmetry transformation of a scalar $\Phi$ is given by its fermionic superpartner $\Psi$ through the Killing spinor $\varepsilon$ as
\begin{align} \label{eq:Qchiral}
\begin{split}
Q \Phi = {{\ve}} \Psi \,. 
\end{split}
\end{align}
If the scalar and spinor fields are expanded in terms of the corresponding eigenfunctions as
 \begin{align} \label{eq:phiexp}
      \Phi & =  \int_{\mathbb{R}_{\geq 0}} {\rm d}\lambda ~ \sum_{k \in \mathbb{Z}}\varphi(\lambda,\, k) \phi_{\lambda,\, k}\, \\
        \label{eq:psiCexp}   \Psi & =  \int_{\mathbb{R}} {\rm d}\lambda ~ \sum_{q =\pm,\, k \in \mathbb{Z}_{\geq 0}} \theta_{q}(\lambda,\, k) \psi^q _{\lambda\, , k} \, ,
  \end{align}
  then supersymmetry of the expansion coefficient of the scalar is give by 
   \begin{align} \label{eq:Qphi}
      Q \varphi(\lambda,\, k) & =\int_{\mathbb{R}} {\rm d}\lambda^\prime ~ \sum_{q =\pm,\, k^\prime \in \mathbb{Z}_{\geq 0}} \theta_{q}(\lambda^\prime,\, k^\prime) \Big{\langle} \phi_{\lambda, \, k} \, \Big{|} \,   \varepsilon \psi^q _{\lambda^\prime\, , k^\prime} \Big{\rangle}\, . 
  \end{align}

  The inner product in \eqref{eq:Qphi} seems ill-defined due to the asymptotic growth of the Killing spinor~$\varepsilon$. However, we can obtain the meaningful result using analytic continuation. To evaluate the inner product, we note the relations given by
  \begin{align} \label{eq:bispinor0}
 \begin{split}
\epsilon \psi^+_{\lambda,k} & =  \bar \alpha^+_{\l , \, k}  \sc_{\l -\frac{\i}{2}, k+1}   \,, \, \qquad
\epsilon \psi^-_{\lambda,k} = \bar \alpha^-_{\l , \, k} \sc_{\l -\frac{\i}{2}, -k}   \,.
\\
\bar \epsilon \psi^+_{\lambda  ,k} & =\alpha^+_{\l , \, k} \sc_{\l -\frac{\i}{2}, k}   \,, \qquad 
\bar \epsilon \psi^-_{\lambda ,k} = \alpha^-_{\l , \, k} \sc_{\l -\frac{\i}{2}, -k-1}   \, , 
\end{split}
\end{align}
where 
\begin{align} \label{eq:Alphasdef1}
\begin{split}
    \bar \alpha^+_{\l , \, k} & = \alpha^-_{\l , \, k} = - \sqrt{\frac{1+k + \i \l}{  2\i\l+1}}  \,, \quad \quad
    \bar \alpha^-_{\l , \, k}  = \alpha^+_{\l , \, k} = - \sqrt{ \frac{k - \i \l}{ 2\i\l+1}}\,.
    \end{split}
\end{align}
Since the bi-spinors are proportional to the scalar eigenfunctions with shifted spectral parameter, e.g. $\phi_{\l- \i/2 , k}$, the analytically continued result of the inner product in \eqref{eq:Qphi} can be evaluated using
\begin{align}\label{innerSscalar}
 \Big{\langle} \phi_{\lambda, \, k} \, \Big{|} \, \varepsilon \psi^q_{\l',k'}
\Big{\rangle}& \;\sim \; 
\left(\delta (\lambda - \lambda^\prime + \tfrac{\i}{2}) +\delta (\lambda + \lambda^\prime  -\tfrac{\i}{2})\right)  \, .
\end{align}
Here, the fact that the spectral parameters $\l$ for scalar and $\l'$ for spinor are real for the standard eigenfunctions results in vanishing of the inner product, and therefore no superpartner of the scalar coefficient is obtained for the standard eigenfunctions. 

The result of the inner product in \eqref{innerSscalar} immediately suggests complexifying spectral parameter.  The most natural way of complexification is universally shifting the spectral parameter of the spinor field as $\l \rightarrow \l +\i/2$, that is to use $\psi_{\l +\i/2,k}^\pm$ as the basis functions.  This is because it makes all the bi-spinors in  \eqref{eq:bispinor0} proportional to the standard scalar basis $\phi_{\l,k}$ and thus \eqref{innerSscalar} becomes Dirac delta function with real parameters making the supersymmetry relation manifest. Also, although the asymptotic behaviour of the spinor basis is changed from \eqref{Fermionasympt} to
\be\label{AsympSpinor}
\psi^\pm_{\l +\frac{\i}{2},k} \sim  {\rm e}^{\pm \i(k+\frac{1}{2})\theta}   \Biggl( \pm {\rm e}^{-\i \lambda \eta} \wt{B}_{-\lambda -\frac{\i}{2},k}  \biggl( \ba{c} 1 \\  1  \ea \biggr)+ {\rm e}^{- \eta}{\rm e}^{\i \lambda \eta} \wt{B}_{\lambda+\frac{\i}{2},k}  \biggl( \ba{c} 1 \\  -1  \ea \biggr) \Biggr) ,
\ee
having degree of growth $0$, 
the orthonormality of spinor fields as in \eqref{innerprodfermion} is preserved as
\beqa\label{orthonormalfermionNew}
\langle \psi^\pm_{\lambda + \frac{\i}{2}\, k} | \psi^\pm_{\lambda' + \frac{\i}{2}\, k'} \rangle 
& \equiv & \pm \i \int {\rm d}\eta {\rm d}\theta \sqrt{g}\, \psi^\mp_{\lambda + \frac{\i}{2},k} \psi^\pm_{\lambda' +\frac{\i}{2},k'} 
= \delta(\lambda - \lambda')\delta_{k,k'}\,,
\eeqa
without relying on any analytic continuation. The leading behavior in \eqref{AsympSpinor} is canceled in the inner product \eqref{orthonormalfermionNew}. For the proof, we refer Appendix \ref{appendixB}.

One may try to consider another way of complexification of the spectrum. For instance, suppose we consider the general way of shifting $\l\rightarrow \l + \i x$ with $-1/2 \leq x <1/2 $ for the spectrum  of spinor field. Then \eqref{innerSscalar} implies that one has to consider $\phi_{\l + \i (x -1/2)}$ as well as $\phi_{\l -\i (x -1/2)}$  for the scalar  basis. This is problematic because the degrees of freedom of the scalar seems to have doubled, and furthermore we do not find a good definition of inner product rendering these eigenfunctions normalizable. 

\subsubsection{Supersymmetry of vector} \label{SUSYVec}
The problem for the supersymmetry transformation of vector field is more subtle because it involves the gamma matrix.  The generic form of the transformation is given by 
\begin{eqnarray}\label{Qvector}
    Q A_{\mu} = \varepsilon \gamma_{\mu} \Psi\,. 
\end{eqnarray}
Together with the mode expansion of the spinor field as in \eqref{eq:psiCexp}, the vector field has the following mode expansion using the basis functions \eqref{eq:AdS2VecEigenfunction} and \eqref{nonnormalizablescalarsads2}:
  \begin{align} 
     A_{\mu} & = \int_{\mathbb{R}_{\geq 0}} {\rm d}\lambda ~ \sum_{k \in \mathbb{Z}} \left( a_{1}(\lambda,\, k) U^{(1)}_{\mu,\l,k} 
     +a_{2}(\lambda,\, k) U^{(2)}_{\mu,\l,k}
     \right) + \sum_{\ell \, \in \, \mathbb{Z}- \{0\}} a_{0(\ell)} \partial_{\mu} \Lambda_{(\ell)} \,  .  \label{eq:Aexp0} 
  \end{align} 
To find the supersymmetry transformation of the expansion coefficient, we have to consider the following inner products
 \begin{align} \label{eq:innerQalk0}
\begin{split}
    Q a_{a}(\l, \, k) & = \int_{\mathbb{R}} {\rm d}\lambda^\prime ~ \sum_{q =\pm,\, k^\prime \in \mathbb{Z}_{\geq 0}} \theta_{q}(\lambda^\prime,\, k^\prime)\Big{\langle}U^{(a)\mu}{}_{\lambda,\, k} \Big{|}\,   \varepsilon \gamma_{\mu} \psi^q_{\l'\,, k'} \Big{\rangle} \,,
\\
Q a_{0(\ell)}  &=  \int_{\mathbb{R}} {\rm d}\lambda^\prime ~ \sum_{q =\pm,\, k^\prime \in \mathbb{Z}_{\geq 0}} \theta_{q}(\lambda^\prime,\, k^\prime)\Big{\langle}\partial^\mu \Lambda_{(\ell)} \Big{|}\, \varepsilon \gamma_{\mu} \psi^q_{\l'\,, k'}\Big{\rangle} \,. 
    \end{split}
\end{align} 

Although the inner product seems ill-defined due to the asymptotic growth of the Killing spinor $\varepsilon$, we can obtain the meaningful result using analytic continuation.  For the evaluation, we find the relation between the bifermions and the basis functions of the vector fields. For $\l \neq 0$, let us first consider  splitting the bifermions as  
\begin{align}\label{Splitingbiferm}
    \begin{split}
         \epsilon \gamma_\mu \psi^\pm_{\l \, , k} = -  \frac{1}{\i \l} \partial_\mu (\epsilon \psi^\pm_{\l \, , k} ) +\overline{\Psi}^\pm_{\mu\,,\l,k}\,,&\qquad \overline{\Psi}^\pm_{\mu\,,\l,k}\equiv \left(\epsilon \gamma_\mu\psi^\pm_{\l \, , k}   +  \frac{1}{\i \l} \partial_\mu (\epsilon \psi^\pm_{\l \, , k} )\right) \,,
   \\
      \bar\epsilon \gamma_\mu \psi^\pm_{\l \, , k} =  -\frac{1}{\i \l} \partial_\mu (\bar\epsilon \psi^\pm_{\l \, , k} ) +{\Psi}^\pm_{\mu\,,\l,k}\,,&\qquad {\Psi}^\pm_{\mu\,,\l,k}\equiv \left(\bar\epsilon \gamma_\mu\psi^\pm_{\l \, , k}   +  \frac{1}{\i \l} \partial_\mu (\bar\epsilon \psi^\pm_{\l \, , k} )\right) \,. 
    \end{split}
\end{align}
Then, the first terms in \eqref{Splitingbiferm} have the following relation with the basis function in \eqref{eq:AdS2VecEigenfunction}  
\begin{align} \label{eq:delbifU}
    \begin{split}
        \frac{1}{\sqrt{\k_{\l- \frac{\i}{2}}} } \partial_\mu (\epsilon \psi^+_{\l,k}) &=  \bar \a_{\l, \, k}^+ U^{(1)}_{\mu, \l-\frac{\i}{2} \,,k+1}\,,\qquad     \frac{1}{\sqrt{\k_{\l- \frac{\i}{2}}} }\partial_\mu (\epsilon \psi^-_{\l,k})\= \bar \a_{\l, \, k}^- U^{(1)}_{\mu, \l-\frac{\i}{2} \,,-k}\,,
        \\
          \frac{1}{\sqrt{\k_{\l- \frac{\i}{2}}} }  \partial_\mu (\bar\epsilon \psi^+_{\l,k}) &= \alpha^+_{\l, \, k} U^{(1)}_{\mu, \l-\frac{\i}{2} \,,k}\,,\qquad    \frac{1}{\sqrt{\k_{\l- \frac{\i}{2}}} } \partial_\mu (\bar\epsilon \psi^-_{\l,k})\= \alpha^{-}_{\l, \, k} U^{(1)}_{\mu, \l- \frac{\i}{2}\,,-k-1}\,,
    \end{split}
\end{align}
where the prefactors are given in \eqref{eq:Alphasdef1}.
The second terms in \eqref{Splitingbiferm}, namely $\Psi^\pm_{\mu}(\l,k)$ and $\overline{\Psi}^\pm_{\mu}(\l,k)$ have the following relation with the basis functions in \eqref{eq:AdS2VecEigenfunction} 
\begin{align}
\begin{split} \label{eq:Psimu}
  \bar{\Psi}_{\mu}^{+}(\lambda,\, k) & = \bar \beta^+_{\l , \, k} U^{(2)} _{\mu,\l +\frac{\i}{2},k+1}\, , \quad \quad 
    \bar{\Psi}_{\mu}^{-}(\lambda , \, k)  =  \bar \beta^-_{\l , \, k} U^{(2)} _{\mu,\l+\frac{\i}{2},-k}\, , \\
       \Psi_{\mu}^{+}(\lambda , \, k) & =   \beta^+_{\l ,\, k} U^{(2)} _{\mu,\l+\frac{\i}{2},k}\, , \quad \quad   
      \Psi_{\mu}^{-}(\lambda, \, k)  =      \beta^-_{\l , \, k} U^{(2)} _{\mu,\l+\frac{\i}{2},-k-1}\, ,
      \end{split}
  \end{align} 
  where
  \begin{align} \label{eq:Betasdef1}
      \bar \beta^+_{\l , \, k} & = - \beta^-_{\l ,\, k}\equiv - \frac{\sqrt{\k_{\l+\frac{\i}{2}}}}{\i\lambda}\sqrt{\frac{1+k  -\i \l}{2\i \l - 1}}\, ,  \qquad
      \bar \beta^-_{\l , \, k}  = - \beta^+_{\l , \, k} \equiv  \frac{\sqrt{\k_{\l+\frac{\i}{2}}}}{\i\lambda}\sqrt{\frac{ k +\i \l}{2\i \l - 1}} \,.
  \end{align}
For $\l = 0$, the splitting in \eqref{Splitingbiferm} is singular. Therefore, we instead note the following relations with the basis for boundary zero modes of vector fields \eqref{nonnormalizablescalarsads2} as 
\begin{align}\label{BifermZero}
    \begin{split}
       \epsilon \gamma_\mu \psi^+_{\l,k} &= -\i \sqrt{2\pi (k+1)}\partial_\mu \Lambda_{k+1}\,\delta(\l) \,,~\quad
\epsilon \gamma_\mu \psi^-_{\l,k} = -\i \sqrt{2\pi k}\,\partial_\mu \Lambda_{-k}\delta(\l)\,, \\
\bar{\epsilon}\gamma_\mu \psi^+_{\l,k} &= -\i \sqrt{2\pi k}\,\partial_\mu \Lambda_{k}\delta(\l)\,, \quad\qquad\qquad
\bar{\epsilon}\gamma_\mu \psi^-_{\l,k} = -\i \sqrt{2\pi (k+1)}\,\partial_\mu \Lambda_{-k-1}\delta(\l)\,. 
    \end{split}
\end{align}
Here, we note that $\bar{\epsilon}\gamma_\mu \psi^+_{\l,0}$ and $\epsilon \gamma_\mu \psi^-_{\l,0}$ are zero.

Using the relation between the bifermions and the bosonic basis as from \eqref{Splitingbiferm} to \eqref{BifermZero}, we can find the analytic continued result of the inner product in \eqref{eq:Aexp0} to find the superpartners of the expansion coefficients, $Q a_{1,2}$ and $Q a_{0(\ell)}$. The results are schematically given as
\begin{subequations}\label{InnerVF}
\begin{align}
\label{eq:deltaminus}
    \Big{\langle}  U^{(1)}_{\mu,\lambda,\, k}\, \Big{|} \,\varepsilon\gamma^{\mu}  \psi_{\lambda^\prime, k^\prime} \Big{\rangle} & \sim \delta (\lambda - \lambda^\prime + \tfrac{\i}{2}) +\delta (\lambda + \lambda^\prime - \tfrac{\i}{2})  \,, \\
\label{eq:deltaplus}
    \Big{\langle} U^{(2)}_{\mu,\lambda,\, k}\, \Big{|} \,\varepsilon\gamma^{\mu}  \psi_{\lambda^\prime, k^\prime}\Big{\rangle}  &\sim  \delta (\lambda - \lambda^\prime - \tfrac{\i}{2}) +\delta (\lambda + \lambda^\prime  +\tfrac{\i}{2})  \, ,\\
    \label{innerZMbiSpinor}
    \Big{\langle} \partial_{\mu}\Lambda_{(\ell)}\, \Big{|} \,\varepsilon\gamma^{\mu}  \psi_{\lambda^\prime, k^\prime}\Big{\rangle}  &\sim  \delta (\lambda^\prime )  \, .
\end{align}
\end{subequations}
Here, the fact that the standard spectral parameter $\l$ for the vector and $\l'$ for the spinor are real results in the vanishing of the inner products \eqref{eq:deltaminus} and  \eqref{eq:deltaplus}. Crucially the inner product \eqref{innerZMbiSpinor} does not vanish for real $\l^\prime$. Therefore, there is no superpartner of the expansion coefficient of continuous spectrum, but there is superpartner of the expansion coefficient of discrete boundary zero modes.

\begin{figure} 
\centering
	\begin{tikzpicture}[scale=1.15]


	\draw[help lines,->] (-4-7.5,0) -- (2-7.5,0) coordinate (xaxisB);
		
	\draw[help lines,->] (-1-7.5,-1.5) -- (-1-7.5,2.5) coordinate (yaxisB);
		
		  	\node [below] at (xaxisB) {$z_R$};
			\node [above] at (yaxisB) {$ z_I$};
		
		   	\node at (-1-7.5,2) {{\color{red}$\times$}};
		   	 	\node at (-1-7.5,-1) {{\color{red}$\times$}};
			
			\node [font=\Large,left] at (-1-7.5,2) {$2 \, \i \hspace{2mm}$};

   	\node at (-1-7.5,1) {{\color{red}$\times$}};

			\node [font=\Large,left] at (-1-7.5,1) {$\i \hspace{2mm}$};
				\node [font=\Large,left] at (-1-7.5,-1) {$- \,\i \hspace{2mm}$};
		
		
		

		\draw[thick] (-4-7.5,0.5) --(0.5-7.5, 0.5);
	    \draw[thick, ->] (0.5-7.5,0.5) --(0.6-7.5, 0.5);
	     \draw[thick] (0.5-7.5,0.5) --(2-7.5, 0.5);
	    	\node [font=\Large,above] at (0.59-7.5, 0.6) {$\psi_{\l +\frac{\i}{2}}$};
	\node[circle, draw, fill=white,inner sep=0pt,minimum size=4pt] (c) at (-1-7.5,0.5){};

	\draw[help lines,->] (-4-0.75,0) -- (2-0.75,0) coordinate (xaxisB);
		
	\draw[help lines,->] (-1-0.75,-1.5) -- (-1-0.75,2.5) coordinate (yaxisB);
		
		  	\node [below] at (2.0-0.75,0) {$z_R$};
			\node [above] at (yaxisB) {$z_I$};
		
		   	\node at (-1-0.75,2) {{\color{red}$\times$}};
		   	 	\node at (-1-0.75,-1) {{\color{red}$\times$}};
			
			\node [font=\Large,left] at (-1-0.75,2) {$2 \, \i \hspace{2mm}$};

   	\node at (-1-0.75,1) {{\color{red}$\times$}};

			\node [font=\Large,left] at (-1-0.75,1) {$\i \hspace{2mm}$};
				\node [font=\Large,left] at (-1-0.75,-1) {$- \,\i \hspace{2mm}$};

		\draw[thick, ->] (-4-0.75,-0.5) --(-2.5-0.75,- 0.5);
  \draw[thick] (-2.5-0.75,-0.5) --(-1-0.75,- 0.5);
  		\draw[ thick, ->] (-4-0.75,0.5) --(-2.5-0.75, 0.5);
    \draw[ thick] (-2.5-0.75,0.5) --(-1-0.75, 0.5);
	    \draw[thick, ->] (-1-0.75,0.5) --(0.6-0.75, 0.5);
	     \draw[thick] (0.5-0.75,0.5) --(2-0.75, 0.5);
      \draw[ thick, ->] (-1-0.75,-0.5) --(0.6-0.75, -0.5);
	     \draw[thick] (0.5-0.75,-0.5) --(2-0.75, -0.5);
	    	\node [font=\Large,above] at (0.59-0.75, 0.6) {$\chi_{\l+\frac{\i}{2}}$};
      \node [font=\Large,below] at (0.59-0.75, -0.6) {$\chi_{\l-\frac{\i}{2}}$};
     
	\node[circle, draw, fill=black,inner sep=0pt,minimum size=4pt] (c) at (-1-0.75,0){};
 	\node [font=\Large,above] at (-0.5-0.75,0) {$\chi_{_0}$};

	\end{tikzpicture}

  \caption{ The figure shows the complex plane $z = z_R + \i z_I$ associated to the spectral parameter of the fermion in the chiral multiplet (left side) and the vector multiplet (right side). In each case we draw the two contours associated to the spectral parameter corresponding to eigenfunctions $\psi^q_{\l +\frac{\i}{2}, \, k}$ and  $\psi^q_{\l -\frac{\i}{2}, \, k}$ ($\lambda \in \mathbb{R}$). With red crosses we indicate the points where the spectral density \eqref{spectralF} have poles. On the left-hand side, the origin $z=0$ corresponds to the discrete set of superpartners $\chi_0$ of boundary zero modes for the vector field.}
		\label{fig:theonlyfigure}
\end{figure}
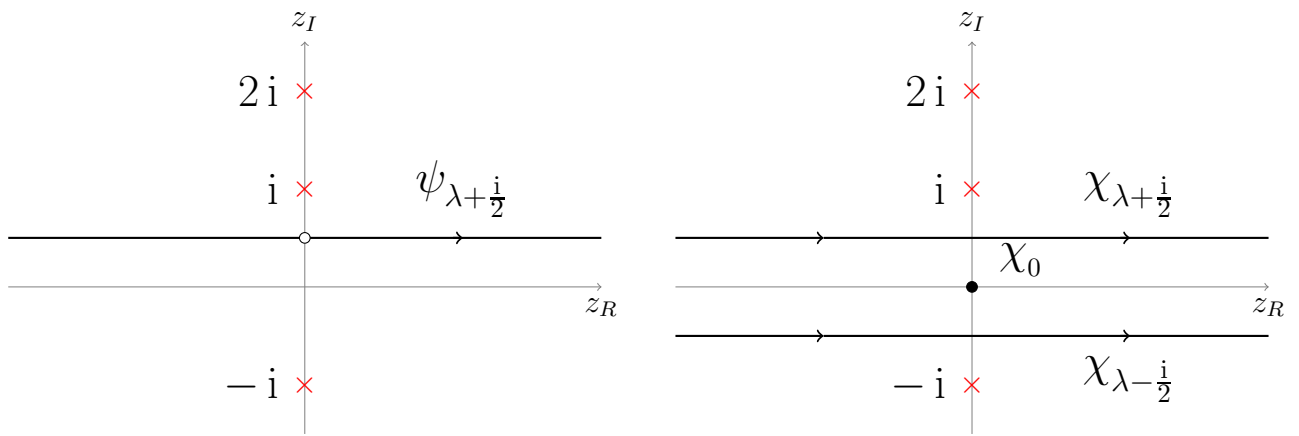

Similar to the case for the supersymmetry of scalar, the results of the inner product \eqref{InnerVF} suggest to consider complexified spectrum. However, the structure is different. First of all, the result \eqref{innerZMbiSpinor} implies that there is an isolated point in the spectrum of gaugini, which is at $\lambda=0$, such that it is superpartner of the boundary zero modes of vector field. The corresponding eigenfunction is normalizable, and thus this  spectrum should be in consideration. Secondly, for the $\l \neq 0$, we can consider the shift of $\lambda$ by $\i/2$, but universal shift either of $\lambda +\i/2 $ or $\lambda -\i/2$ does not work because the \eqref{eq:deltaminus} and \eqref{eq:deltaplus} suggest the opposite shift of $\lambda$. If we have two spinor fields, then we can take one sector having $\psi_{\l +\i/2}^\pm$ such that \eqref{eq:deltaminus} is non-zero, and the other sector having    $\psi_{\l -\i/2}^\pm$ such that \eqref{eq:deltaplus} is non-zero. 
Since the vector field has two different sectors of continuous spectrum by $U^{(1)}_{\mu,\l,k}$ and $U^{(2)}_{\mu,\l,k}$, it is plausible that we need two gaugini for them to reflect corresponding two sectors.
The schematic description of the complexified spectrum of gaugini is depicted in Figure \ref{fig:theonlyfigure}, in comparison to the spectrum of spinor that is superpartner of scalar field. In the following section we present the details about the supersymmetry relation among expansion coefficient of boson and fermions in chiral and vector multiplets.

Note that, we could have considered other possibilities of complex spectrum: An example is to implement a universal shift of the gaugini spectrum either by $\lambda+ \i/2 $ or $\lambda- \i/2 $. However, in this case, we have to consider both the $U^{(1)}_{\lambda + \i }$ and $U^{(1)}_{\lambda - \i }$ or both the  $U^{(2)}_{\lambda + \i}$ and $U^{(2)}_{\lambda - \i}$ for the spectrum of vector field, and we do not find a good way of treat them in a normalizable manner. 
Therefore, we will not consider the complex shift of spectrum for vector field, keeping their basis functions normalizable. 
This seems to fit with the physical intuition: The fact that the \eads2 partition function is well-defined only for charge-fixing boundary condition \cite{Sen:2008vm} requires us to use normalizable condition for the dynamical modes, i.e. $U^{(2)}_{\l}$ with real $\lambda$. Moreover, from the definition of pure gauge mode, that is with normalizable parameters, we also impose the dynamical mode normalizable, i.e. $U^{(1)}_{\lambda}$ with real $\lambda$.  


\section{Supersymmetric spectrum}\label{sec:Resolutionoftheproblems}
In this section, we elaborate on the idea suggested in section \ref{sec:Problemsofsupersymmetry} to find the supersymmetric spectrum taking examples of  $\cN=(2,2)$  supermultiplets on \eads2. Specifically, we identify the superpartners of the expansion coefficients of bosonic and fermionic fields   related by the Killing spinors in \eqref{eq:KillingSpinors}.\footnote{They  are $\cN=(1,1)$ subsector of the full set of supercharges. The extension of this analysis to include the remaining Killing spinors is straightforward, but they do not provide additional conceptual clarity. }
We begin by considering a chiral multiplet in section \ref{subsec:chiralmul} and move on to the vector multiplet in section~\ref{sec:vecmultiplet}. 
\subsection{Chiral multiplet} \label{subsec:chiralmul}
The chiral multiplet consists of $4+4$ bosonic and fermionic degrees of freedom, namely $\{ \phi\,, \overline{\phi}\,, \psi\,,\overline{\psi}\,, F\,, \overline{F}\}$, and the supersymmetry transformation rule is given in  \eqref{eq:Qchiral01}. The supersymmetric structure is more manifest if we organize the variables into the cohomological form by rearranging the fermionic fields as $\{\bar\epsilon \psi\,,\epsilon\bar\psi\,, \epsilon \psi\,, \bar\epsilon \bar\psi\}$.  As can be seen from \eqref{eq:ChBPS3}, this is because the field content can be grouped into a set of $2+2$ degrees of freedom which we call ``elementary variables'',  $\{ \phi\,, \bar\phi\,;\, \epsilon \psi\,, \bar\epsilon \bar\psi\}$, and their corresponding superpartners. Therefore, it is convenient to find the supersymmetric spectrum and their supersymmetry relation using this cohomological arrangement. For more detail about the cohomological variables we refer to Appendix~\ref{app:twistedvariable}.

As was suggested in section \ref{SUSYScalar}, we use the basis function $\{ \phi_{\l,k} \,|\,\l \in \mathbb{R}_{>0}\,, k\in \mathbb{Z}\}$ both for the scalar fields, $\phi\,, \overline{\phi}$, and for the auxiliary scalars, $F,\overline{F}$, and $\{\psi^\pm_{\l+\i/2 ,k}\,| \, \l \in \mathbb{R}\,, k \in\mathbb{Z}_{\geq 0} \}$ for the spinor fields, $\psi, \overline{\psi}$. Note again that these basis functions are delta-function normalizable with respect to the corresponding definition of inner product. Let the corresponding expansion coefficients with respect to the basis functions be $\{ \varphi(\l,k)\,, \overline{\varphi}(\l, k)\,, \theta_\pm(\l+ \i/2,k)\,, \bar{\theta}_\pm(\l+\i/2 ,k)\,, f(\l,k)\,, \bar{f}(\l,k)\}$. Then, by inserting the expansion of fields into the supersymmetry relation \eqref{eq:Qchiral01}, we obtain the superpartners of the $\varphi_{\l,k}$ with $\l \geq 0$ and $k\in \mathbb{Z}$  as 
\be
Q\varphi_{\l,k}\= \begin{cases}
  \alpha^+_{\l + \frac{\i}{2},k}\theta_{+}(\l+\frac{\i}{2},k)+ \left(\l \rightarrow - \l\right)  \,,& \mbox{for }k\geq 0\,, \vspace{3mm}
    \\
     \alpha^-_{\l + \frac{\i}{2} , \, |k|-1} \theta_{-}({\lambda + \frac{\i}{2},\, |k|-1}) + \left(\l \rightarrow - \l\right)  \,,& \mbox{for }k< 0\,,
\end{cases}
\ee
and similarly obtain superpartners of $\overline{\varphi}_{\l,k}$ as
\be
Q\bar\varphi_{\l,k}\= \begin{cases}
  \bar\alpha^+_{\l + \frac{\i}{2},k}\bar\theta_{+}(\l+\frac{\i}{2},k)+ \left(\l \rightarrow - \l\right)  \,,& \mbox{for }k\geq 0\,, \vspace{3mm}
    \\
    \bar\alpha^-_{\l + \frac{\i}{2} , \, |k|-1} \bar\theta_{-}({\lambda + \frac{\i}{2},\, |k|-1}) + \left(\l \rightarrow - \l\right)  \,,& \mbox{for }k< 0\,,
\end{cases}
\ee
where $\alpha^{\pm}_{\l+ \i/2,k}$ and $\bar{\alpha}^\pm_{\l+\i/2,k}$ are the constants given in \eqref{eq:Alphasdef1}.

We can obtain the supersymmetry relation between the auxiliary scalars and the spinor fields by looking at the variation of elementary fermions $\epsilon \psi$ and $\bar{\epsilon}\bar{\psi}$ in cohomological reorganization \eqref{eq:ChBPS3}. By inserting the mode expansion into the supersymmetry transformation of~$\epsilon \psi$, we find that, with $\l \geq 0\,, k\geq 0$,
\begin{eqnarray}
Q\left(\bar{\alpha}^+_{\l+\frac{\i}{2}, k}  \theta_+(\l + \tfrac{\i}{2},  k)+ \left(\l \rightarrow - \l\right)   \right) = - \i  f_{\l, k +1}    - \bigl((k +\tfrac{\i}{2})^2 +\l^2\bigr)^{\frac{1}{2}}\varphi_{\lambda,\, k}  \,,   
    \end{eqnarray}
    \begin{eqnarray}
  Q \left(  \bar{\alpha}^-_{\l+ \frac{\i}{2},  k}\theta_{-}(\lambda + \tfrac{\i}{2}, k) + \left(\l \rightarrow - \l\right)  \right)\!  = \! - \i  f_{\l, -k}    + \! \bigl((k +\tfrac{\i}{2})^2 +\l^2\bigr)^{\frac{1}{2}}\varphi_{\lambda, -k -1}    \, ,     
    \end{eqnarray}
and from the supersymmetry transformation of $\bar\epsilon \overline{\psi}$, we find that, with $\l \geq 0$ and $k\geq 0$
\begin{eqnarray}
Q\left(\alpha^+_{\l+\frac{\i}{2},  k}  \bar\theta_+(\l + \tfrac{\i}{2}, k) + \left(\l \rightarrow - \l\right)   \right)  = - \i  \bar f_{\l, k}    +  \bigl((k +\tfrac{\i}{2})^2 +\l^2\bigr)^{\frac{1}{2}}\bar\varphi_{\lambda,\, k+1}  \,,
    \end{eqnarray}
    \begin{eqnarray}
      \!\!    Q \left(  \alpha^-_{\l+ \frac{\i}{2},  k}\bar\theta_{-}(\lambda + \tfrac{\i}{2}, k) + \left(\l \rightarrow - \l\right)   \right)\!  = \! - \i  \bar f_{\l, -k-1}     - \!\bigl((k +\tfrac{\i}{2})^2 +\l^2\bigr)^{\frac{1}{2}}\bar\varphi_{\lambda, -k } \,.
    \end{eqnarray}
This provides the complete supersymmetric pairing between boson and fermion in the chiral multiplet.
\subsection{Vector multiplet} \label{sec:vecmultiplet}
The vector multiplet consists of $6+6$ degrees of freedom, $5+4$  of which are in the form $\{A_\mu, \rho, \sigma, \chi, \overline{\chi}, D\}$ and $1+2$ in the ghost multiplet $\{c\,, \bar{c}\,, b\}$. Unlike for the chiral multiplet,  the universal shift of the spectral parameter for spinor fields in the vector multiplet does  not recover the supersymmetric pairing as we have discussed in section \ref{SUSYVec}. The main technical reason for this difference 
is that supersymmetry transformations of each bosonic field involve different gamma matrices. As depicted in figure \ref{fig:theonlyfigure}, the gaugini spectrum has three distinct sectors thus providing the superpartner of three corresponding sectors of the vector field. 

The structure of the fermionic spectrum will in turn affect the scalar fields in the vector multiplet by their supersymmetry transformation.
Therefore, we examine in detail the structure of the supersymmetric spectrum for each field and find the supersymmetry relation among the expansion coefficients. 
Once again it will be convenient to rearrange the fields into the cohomological form, that consists of $3+3$  bosonic and fermionic elementary variables $\{ A_\mu\,, \rho\, ; \chi_4\,, c\,, \bar{c}\}$ and their superpartners, where $\chi_4 = - \i (\bar\epsilon \chi + \epsilon \bar{\chi})/2 $ is defined as a reorganization of the gaugini. Their definitions and some of their properties are summarized in the Appendix \eqref{app:twistedvariable}. 

\subsection*{Superpartner of $A_\mu$}
First of all, we begin by considering supersymmetric transformation of vector field
\be\label{susyvector}
Q A_\mu \= -\frac{\i}{2}(\bar\epsilon \gamma_\mu \chi + \epsilon \gamma_\mu \bar\chi)
\ee
and we find the superpartner of the expansion coefficient $a_{i}(\l,k)\,,i=1,2 $ and $a_{0(\ell)}$\,  of the vector field $A_\mu$, whose expansion is shown in \eqref{eq:Aexp0}, i.e., we find $Q a_{i}(\l,k)$ and $Q a_{0(\ell)}$.

As explained after \eqref{InnerVF} and depicted in Figure \ref{fig:theonlyfigure}, the spectrum of gaugini needs two continuous sectors with opposite complex shift as well as an additional isolated point. The two different continuous sectors can be realized since we have two gaugini $\chi$ and $\overline{\chi}$ whose expansion coefficients will be appropriately related. To find such relation, we introduce a doubling of expansion coefficients $\omega_{i}\,, i=1,2$  for $\chi$ and $\overline{\omega}_{i}\,, i=1,2$  for $\overline{\chi}$, where $i=1$ corresponds to the spectrum of $\l+\i/2$ and the $i=2$ to the spectrum of $\l-\i/2$. Then we will require linear relations between $\omega_i$ and $\overline{\omega}_i$ such that only $\omega_1$ is related to $Qa_1(\l,k)$ and only $\omega_2$ is related to $Qa_2(\l,k)$.
Taking all this into account, we propose the following mode expansion of the gaugini $\chi$ and~$\bar \chi$: 
\begin{subequations}
\label{eq:fermionintermsofbasis}
\begin{align} \label{eq:fermionintermsofbasisI}
    \chi & =  \sum_{k  \geq 0, \, q=\pm} \int_{\mathbb{R}_\varepsilon} {\rm d} \l \left( \omega_{1,q}(\l ,k ) \psi^{q}_{\l +\frac{\i}{2} , \, k} + \omega_{2,q}(\l , \, k ) \psi^{q}_{\l - \frac{\i}{2}  , \, k }\right) +\sum_{k \geq 0 ,\,  q=\pm} \int_{-\varepsilon}^{\varepsilon} {\rm d} \l ~ \omega_{q}(\l , k)  \psi^{q}_{\l , \, k} \\
    \bar \chi  & = \sum_{k \geq 0  , \, q=\pm} \int_{\mathbb{R}_\varepsilon} {\rm d} \l  \left( \bar \omega_{1,q}(\l ,k ) \psi^{q}_{\l + \frac{\i}{2}, \, k} + \bar \omega_{2,q}( \l, \, k ) \psi^{q}_{\l - \frac{\i}{2} , \, k }\right) +\sum_{k  \geq 0 ,\,    q=\pm} \int_{-\varepsilon}^{\varepsilon} {\rm d} \l ~ \bar{\omega}_{q}(\l , k) \psi^{q}_{\l , \, k}\,, \label{eq:fermionintermsofbasisII}
\end{align}
\end{subequations}
where $\mathbb{R}_{\varepsilon} \equiv \mathbb{R} - (- \varepsilon, \varepsilon)$. To represent the isolated point in the spectrum,  we have introduced the tunable parameter $\varepsilon$ in the last terms of \eqref{eq:fermionintermsofbasis} that we shall later take to zero. Although the $\l \in (- \varepsilon, \, \varepsilon)$  sectors of the spectrum seems to contribute trivially to the fermionic expansion as we take $\varepsilon\rightarrow 0$, it is clear from the multiplication of the Killing spinors and the gaugini \eqref{BifermZero} that, upon inserting their contribution into the right-hand side of the supersymmetry relation in \eqref{susyvector}, these terms account for the superpartners of boundary zero modes. Thus, we find the supersymmetry relation of the boundary zero modes to be 
\begin{equation}
     \label{eq:CzmL1}
    Q a_{0(\ell)}  =\begin{cases}
        - \sqrt{\frac{\pi \ell}{2}} \left(\omega_{+}(0,\, \ell)  + \bar{\omega}_{+}(0,\ell -1 )\right)\, \vspace{1.5mm} & \mbox{for } \ell >0\,,\\ 
         - \sqrt{\frac{\pi |\ell|}{2}} \left(\omega_{-}(0,\, |\ell|-1)  + \bar{\omega}_{-}(0,|\ell| )\right)\, & \mbox{for } \ell <0\,.
    \end{cases} 
\end{equation}
Note that $\omega_{+}(0,0)$ and $\bar\omega_{-}(0,0)$ are not involved in the supersymmetry relation because the bi-spinors with corresponding spinor basis, $\bar\epsilon \gamma_\mu \psi^+_{0,0}$ and $\epsilon \gamma_\mu \psi^-_{0,0}$ are zero as given in~\eqref{BifermZero}.

Now, we want to find the linear relation between  $\omega_i$ and $\overline{\omega}_i$ such that
\begin{equation}
    Q a_{1} \sim \omega_{1,\pm}\,,\qquad Q a_{2} \sim \omega_{2,\pm}\,,
\end{equation}
i.e. $\omega_{2}$ and $\overline\omega_{2}$ do not appear in $Q a_1$, and $\omega_{1}$ and $\overline\omega_{1}$ do not appear in $Q a_2$. Note that inserting the mode expansion \eqref{eq:fermionintermsofbasis}  into the right-hand side of \eqref{susyvector} shows that the coefficient $\omega_{1,\pm}$ and $\bar{\omega}_{1, \pm}$ involve $U^{(1)}_{\l,k}$ and $U^{(2)}_{\l+\i,k}$ sectors, and the coefficients $\omega_{2, \pm}$ and $\bar{\omega}_{2, \pm}$ involve $U^{(1)}_{\l+\i,k}$ and $U^{(2)}_{\l,k}$ sectors. Therefore, as the $U^{(a)}_{\l +\i,k}$ are non-normalizable, it is natural to require eliminating the non-normalizable sector. This is attained by imposing the following relations for $k \geq 0$
\begin{align} \label{eq:constraintA}
\begin{split}
\textrm{for $k>0$},\quad 0 &= \omega_{2,+}(\l, \, k) \a^+_{\l - \frac{\i}{2}, \, k} + \bar \omega_{2,+}(\l,\, k-1) \bar \a^+_{\l - \frac{\i}{2}, \, k-1}  \,, \\
 0&= \omega_{2,-}(\l, \, k-1) \a^-_{\l - \frac{\i}{2}, \,k-1} + \bar \omega_{2,-}(\l,\, k) \bar \a^-_{\l - \frac{\i}{2}, \, k} \, ,
  \\
    \textrm{for $k=0$},\quad  0&=   \omega_{2,+}({\l , 0}) \a^+_{\l -\frac{\i}{2}, 0} +   \bar \omega_{2,-}({\l , 0}) \bar \a^-_{\l -\frac{\i}{2}, 0}  \, ,
    \\
\textrm{similarly:}\quad  0&=(\omega_{2,\pm}\, \rightarrow \omega_{1,\pm})\quad \& \quad    (\alpha^\pm_{\l -\frac{\i}{2}} \rightarrow \b^\pm_{\l +\frac{\i}{2} })\,.
\end{split}
\end{align} 
Therefore, inserting the mode expansions  \eqref{eq:Aexp0} and  \eqref{eq:fermionintermsofbasis}  into \eqref{susyvector} and using the relations~\eqref{eq:constraintA}, we find the supersymmetry relation between the bosonic and fermionic expansion coefficients. The upshot of this procedure  is that, for $k \in \mathbb{Z}$ and $\l \geq \varepsilon$, we have
\begin{align} 
     Q a_1(\l, k) \label{eq:Qbkp00} &   =  \begin{cases}\frac{\sqrt{\k_\l}}{2 \left(\l + \frac{\i}{2}\right)} \left(  \a^+_{\l+\frac{\i}{2}, \, k} - \frac{ \b^+_{\l+\frac{\i}{2},\, k} \bar \a^+_{\l+ \frac{\i}{2}, \, k-1}}{\bar \b^+_{\l + \frac{\i}{2}, \, k-1}} \right)   
     \omega_{1, +}(\l, k)+ \left(\l \rightarrow - \l\right)\, & \text{if} \quad k \geq  0\vspace{2.0mm} \\
         \frac{\sqrt{\k_\l}}{2 \left(\l + \frac{\i}{2}\right)}\left(\a^-_{\l + \frac{\i}{2}, \,-k-1} -  \frac{ \b^-_{\l + \frac{\i}{2},\, -k-1} \bar \a^-_{\l + \frac{\i}{2}, \, -k}}{\bar \b^-_{\l + \frac{\i}{2}, \, -k} }\right)
       \omega_{1, -}(\l, -k-1) \\
       \quad+ \left(\l \rightarrow - \l\right) & \text{if} \quad k<0 \, ,
       \end{cases}
\end{align}

\begin{align} \label{eq:QamupmK0}
   Q a_2(\l, k) &   =\begin{cases}- \frac{\i}{2}\left( \beta^+_{\l - \frac{\i}{2}, \, k} - \frac{ \a^+_{\l - \frac{\i}{2},\, k} \bar \beta^+_{\l - \frac{\i}{2}, \, k-1}}{\bar \a^+_{\l - \frac{\i}{2}, \, k-1} } \right) ~ \omega_{2,+}(\l, \, k)  + \left(\l \rightarrow - \l\right)\, &  \text{if} \quad k \geq 0 \vspace{2mm} \\
    -\frac{\i}{2} \left(\beta^-_{\l-\frac{\i}{2}, \,k-1} -  \frac{ \a^-_{\l - \frac{\i}{2},\, k-1} \bar \beta^-_{\l - \frac{\i}{2}, \, k}}{\bar \a^-_{\l -\frac{\i}{2}, \, k} }\right)~\omega_{2, -}(\l, k-1)  + \left(\l \rightarrow - \l\right) &  \text{if} \quad k<0\,.
   \end{cases}
\end{align} 
\subsection*{Superpartner of $\rho$}
We want to find the supersymmetry relation between the scalar $\rho$ and gaugini in terms of mode expansion by inserting \eqref{eq:fermionintermsofbasis}   into the following relation
 \begin{align} \label{Qrho}
  \begin{split}
Q \rho  &= - \frac{\i}{2} (\bar\epsilon \gamma_3 \chi + \epsilon \gamma_3 \bar\chi) \, . 
\end{split}
  \end{align} 
 We first note that, on the right-hand side of the above supersymmetry relation, the expansion coefficient $\omega_{1}(\l,k)$ and $\overline{\omega}_1(\l,k)$ in the gaugini would involve the non-normalizable functions $\phi_{\l \pm \i,k}$ but they are canceled due to the relation in \eqref{eq:constraintA}. Thus only $\omega_2$ and $\overline\omega_2$ appear with normalizable basis function of scalar $\phi_{\l,k}$.
 Secondly, there is a peculiar aspect due to the presence of the isolated points in the gaugini spectrum given in the last terms of \eqref{eq:fermionintermsofbasis}, because this demands $\rho$ to have the isolated spectrum in order to meet the supersymmetry relation \eqref{Qrho}. Therefore, we shall have the mode expansion of $\rho$ as follows\footnote{We may write the second term in democratic way using both the $\phi_{\l+\i/2,k}$ and $\phi_{\l-\i/2,k}$ using 
 \be
  \int_{0}^{\varepsilon} {\rm d}\lambda ~ \sum_{k \in \mathbb{Z}} \left(\rho_1(\lambda, k) \phi_{\l + \frac{\i}{2}, \, k}\, + \rho_2(\l,k) \phi_{\l - \frac{\i}{2}, \, k}\right)\,.\nn
  \ee
  In this case, we should note that $\rho_1$ and $\rho_2$ are not independent coefficient because
they are redefinition from \eqref{eq:rhoexp} by 
$\rho_1(\l,k) =\rho_1(\l+{\i}/{2},k)  $ and $\rho_2(\l,k) = \rho_1 ({\i}/{2}-\l,k) $, and thus they are equal at  $\l=0$.
},
 \begin{align} \label{eq:rhoexp}
      \rho & = \int_{\varepsilon}^\infty {\rm d}\lambda ~ \sum_{k \in \mathbb{Z}}\scvrho (\lambda,\, k) \phi_{\lambda,\, k} 
      +  \int_{-\varepsilon}^{\varepsilon} {\rm d}\lambda ~ \sum_{k \in \mathbb{Z}} \rho(\tfrac{\i}{2}+\lambda, k) \phi_{\l + \frac{\i}{2}, \, k}   \,,
\end{align}
where the $\l \in (0, \, \varepsilon)$ part of the $\scvrho$ spectrum is such that it matches the $\l \in (- \varepsilon, \, \varepsilon)$ part of the fermionic spectrum and the symmetry $\phi_{\l, \, k} = \phi_{- \l, \, k}$ has been used.
Upon inserting this mode expansion and the expansion of gaugini \eqref{eq:fermionintermsofbasis} into the supersymmetry relation, we find the supersymmetric transformation of the continuous spectrum for $k \in \mathbb{Z}$ is given by
\begin{eqnarray}
   Q \rho (\l, k) = \begin{cases}
        -\frac{\i}{2} \omega_{2,+} (-\l,k) \left(\alpha_{\l+\frac{\i}{2} ,k}^+  - \frac{\alpha_{-\l-\frac{\i}{2} ,k}^+ \bar \alpha_{\l+\frac{\i}{2},k-1}^+}{\bar \alpha_{-\l-\frac{\i}{2},k-1}^+ }\right)+(\l \rightarrow - \l) & \text{if} \,\, k \geq 0\,
        \\
        -\frac{\i}{2} \omega_{2,-} (-\l,|k|-1) \left(\alpha_{\l+\frac{\i}{2} ,|k|-1}^- - \frac{\alpha_{-\l-\frac{\i}{2} ,|k|-1}^-  \bar \alpha_{\l+\frac{\i}{2},|k|}^-}{\bar \alpha_{-\l-\frac{\i}{2},|k|}^-}\right)+(\l \rightarrow - \l) &  \text{if} \,\, k<0\,,
   \end{cases} 
\end{eqnarray} where we note that  $Q \rho(\l, 0) =0$. We also note that only $\omega_2(\l,k)$ are involved. 
The supersymmetric transformation of the isolated spectrum for $k>0$ and $\l \in (-\varepsilon,\varepsilon)$ is given by 
\begin{eqnarray}
   && Q \rho(\tfrac{\i}{2}+\l,k) = -\frac{\i}{2} \left(\alpha_{-\l,k}^+  \omega _+ (\l,k) + \bar \alpha_{-\l,k-1}^+ \bar \omega _+ (\l,k-1) \right)\,, \\
   && Q \rho(\tfrac{\i}{2}+\l,-k) = -\frac{\i}{2} \left( \alpha_{-\l,k-1}^-  \omega _- (\l,k-1)  + \bar \alpha_{-\l,k}^- \bar \omega _- (\l,k) \right)\,.
\end{eqnarray}
As we take $\varepsilon\rightarrow 0$, i.e. at $\l=0$ they are proportional to the $Q a_{0(k)}$ given in \eqref{eq:CzmL1} as 
\be
Q \rho(\tfrac{\i}{2}, k)
= -\frac{\i}{\sqrt{2\pi}} Q a_{0(k)}\,.
\ee
Therefore, assuming that there is no  $Q$ invariant modes, the isolated spectrum in the $\rho$ parametrized by $\rho({\i}/{2},k)$ does not provide new independent degrees than the boundary zero modes of vector field parametrized by $a_{0(\ell)}$. 
Thus, we may redefine $\rho'$ by subtracting the second term of \eqref{eq:rhoexp} to represent an independent degree of freedom of a boson in vector multiplet. 

Note again that the presence of the isolated spectrum is irrelevant to the dynamics because we take $\varepsilon\rightarrow 0$ limit and the contribution of the spectrum is negligible. 
Further, although the function $\phi_{\pm {\i}/{2}, k}$ is non-normalizable, kinetic term of $\rho$ involves the derivative i.e. $\partial_\mu \phi_{\pm {\i}/{2},k}$, and this is normalizable in terms of inner product as a vector. Therefore, this isolated basis functions does not cause divergence in the action. 

\subsection*{Superpartner of $\chi_4$}
The supersymmetric relation between the auxiliary scalar $D$ and the gaugini can be found by looking at the supersymmetric transformation of $\chi_4= - \i (\bar\epsilon \chi + \epsilon \bar{\chi})/2  $,  given by 
\begin{align} \label{eq:Qchi4}
Q \chi_4 & =\i \bar{\epsilon}\epsilon D + \i  \bar\epsilon \gamma_3 \epsilon  \left(-\i F_{12} + \rho \right) + \i \bar\epsilon \gamma_3 \gamma^\mu \epsilon \partial_\mu \rho\\
&= - D  +\cosh\eta (\i F_{12}-\rho) - \sinh\eta \partial_\eta \rho \,.\nn
\end{align}

Let us first consider the left hand side by looking at $\chi_4$. By inserting the mode expansion of gaugini in \eqref{eq:fermionintermsofbasis} into the expression of $\chi_4$, we find that it has the following mode expansion in terms of the scalar basis function $\phi_{\l,k}$, with additional isolated spectrum as
\begin{align}  \label{eq:Chi4expansion}
\begin{split}
    \chi_4 & =  \int_{\varepsilon}^{\infty} {\rm d} \l ~\sum_{k \in \mathbb{Z}} ~ \chi_4(\l, \, k)\phi_{\lambda,\, k}    +  \int_{-\varepsilon}^{\varepsilon} {\rm d}\lambda ~ \sum_{k \in \mathbb{Z}} \chi_{4}(\tfrac{\i}{2}+\lambda, k) \phi_{\l + \frac{\i}{2}, \, k}\,.
    \end{split}
\end{align}
 With $k>0$, the expansion coefficient $\chi_4(\l,k)$ is given by the coefficient $\omega_1 (\l,k)$ of the gaugini expansion as 
\begin{align} \label{eq:chi4modes}
\begin{split}
  \chi_4(\l , \, k) &   =-\frac{\i}{2} \omega_{1,+}(\l , \, k) \left(\a^+_{\l + \frac{\i}{2} , \, k} - \frac{\bar \a^+_{\l +\frac{\i}{2} , \, k-1}\b^+_{\l+ \frac{\i}{2}, \, k}}{\bar \b^{+}_{\l + \frac{\i}{2} , \, k-1}} \right) + \left(\l \rightarrow -\l\right)\,,\\
   \chi_4(\l , \, -k) &  =-\frac{\i}{2
} \omega_{1,-}(\l  , \, k-1) \left(\a^-_{\l+ \frac{\i}{2} , \, k-1}- \frac{\bar \a^-_{\l + \frac{\i}{2}, \, k} \b^-_{\l + \frac{\i}{2} ,\, k-1}}{\bar \b^-_{\l + \frac{\i}{2}, \, k}}\right)+ \left(\l \rightarrow -\l\right) \, ,
   \end{split}
\end{align}
with $\chi_{4}(\l, 0) =0$, which follows from \eqref{eq:constraintA}. We note that the $\omega_2(\l,k)$ sector of the gaugini expansion does not appear by the relation given in \eqref{eq:constraintA}.
The expansion coefficient~$\chi_4(\i/2 +\l ,k)$ in the second term of \eqref{eq:chi4modes} appears due to the presence of isolated spectrum at $\l=0$ of gaugini as in \eqref{eq:fermionintermsofbasis}, and it is given by
\begin{eqnarray}
   &&  \chi_{4}(\tfrac{\i}{2}+\l,k) = -\frac{\i}{2} \left(\alpha_{\l,k}^+  \omega _+ (-\l,k) + \bar \alpha_{\l,k-1}^+ \bar \omega _+ (-\l,k-1) \right)\,, \\
   &&  \chi_{4}(\tfrac{\i}{2} +\l,-k) = -\frac{\i}{2} \left( \alpha_{\l,k-1}^-  \omega _- (-\l,k-1)  + \bar \alpha_{\l,k}^- \bar \omega _- (-\l,k) \right)\,.
\end{eqnarray}
Specifically, at $\l =0$ they reduce to: 
\begin{eqnarray} \label{eq:Qchi4epsilon}
   &&  \chi_{4}(\tfrac{\i}{2},k) = - \frac{\i}{\sqrt{2 \pi}} Q a_{0(k)}
   \,,
\end{eqnarray}
with $k \neq 0$. Thus, \eqref{eq:Qchi4epsilon} does not provide new relation between boson and fermion.

Now, let us look at the right-hand side of the supersymmetry transformation \eqref{eq:Qchi4epsilon}. Here, there is a subtlety: According to \eqref{eq:Chi4expansion} and \eqref{eq:chi4modes}, up to the isolated spectrum given by~$\chi_4(\i/2+\l,k)$, the left-hand side of \eqref{eq:Qchi4} is written in terms of a normalizable scalar basis with coefficient $\omega_{1, \pm}$, whereas the right-hand side is not normalizable because of the appearance of $\cosh\eta$ and $\sinh\eta$ factors in front of $F_{12}$ and $\rho$ fields.  
Consistency of~\eqref{eq:Qchi4} implies that the auxiliary field $D$ must include the non-normalizable modes such that they cancel the non-normalizable contributions from the $F_{12}$ and $\rho$ terms. 
According to the supersymmetry transformation of $D$ in \eqref{eq:deltaA}, and upon insertion of the gaugini expansion, we find that~\eqref{eq:deltaA} is expressed in terms of $\phi_{\l \pm \i, k}$, which confirms that $D$ is not normalizable. However, as we do not know how to treat the non-normalizable function $\phi_{\l \pm \i,k}$, we prescribe how to deal with $D$ field. 
In the point of view of the cohomological arrangement of fields, $D$ is not an elementary variable in the cohomology complex, thus it is natural to redefine the field $D = D' - (...)$ such that $Q\chi_4 = D'$. In this redefinition, it is straightforward to find the expansion as
\be
D' = \int_{\varepsilon}^{\infty} d\l \sum_{k} D'(\l,k)  \phi_{\l,k} + \int_{-\varepsilon}^{\varepsilon} d\l D'(\tfrac{\i}{2}+\l, k) \phi_{\l+ \frac{\i}{2}, \, k } \,.
\ee
Here, similar to the expansion of $\rho$ in \eqref{eq:rhoexp}, the presence of the isolated spectrum in $D'$ is due to the isolated spectrum of gaugini.

With this expansion, we obtain the supersymmetry relation
\begin{align}
Q \chi_4(\l , k) & = D'(\l,k) \\
Q \chi_{4}(\tfrac{\i}{2}) & = D'(\tfrac{\i}{2},k) \, .
\end{align}
We reiterate that the necessity of redefining the $D$ field implies that imposing normalizable condition (Dirichlet condition) to the $D$ is not compatible with supersymmetry. Furthermore, by the relation \eqref{eq:Qchi4epsilon}, we note that $D'(\i/2,k)$ is not a new degree of freedom. Rather it is related to the $a_{0(k)}$.

\subsubsection*{Superpartner  of ghost $c$
}
The remaining scalar $\sigma$ is related by the supersymmetry of ghost field $c$,
\begin{align}\label{Qc}
Q c &= \Sigma - \Sigma_0\,,\\ \label{eq:Sigma}
\Sigma &\equiv   -v^\mu A_\mu +\i \bar{\epsilon}\epsilon \sigma -  \bar\epsilon \gamma_3 \epsilon \rho =  -A_\theta - \sigma - \i \cosh\eta\, \rho \, ,
\end{align} 
where
$\Sigma_0$ is non-normalizable part of the $\Sigma$ to make the right-hand side of \eqref{Qc} normalizable because the ghost field $c$ and thus the left-hand side is normalizable. In our case $\Sigma_0$ is a constant.\footnote{ The constant mode is from the BPS configuration,  $\rho = {\rho_1}({\cosh\eta})^{-1}$ with $\rho_1 =$constant, which was obtained in \cite{GonzalezLezcano:2023cuh} with different convention where the roles of $\sigma$ and $\rho$ would be interchanged.} Like the auxiliary field $D$, the scalar $\sigma$ must be non-normalizable such that the non-normalizable part due to the $\cosh\eta$ factor in front of $\rho$ and $A_\theta$ must be canceled in the right-hand side of \eqref{eq:Sigma}. 
The supersymmetry transformation of $\sigma$ given in \eqref{eq:deltaA}, with insertion of the gaugini expansion \eqref{eq:fermionintermsofbasis}, would be expressed in terms of the non-normalizable function $\phi_{\l\pm \i,k}$. However,
we do not have a well defined inner product free of divergences. We can avoid directly dealing with an explicit mode expansion for the field $\sigma$ by redefine it as $\sigma = -\Sigma'  -A_\theta - \i \cosh\eta\, \rho - \Sigma_0$, such that $Q c = \Sigma'$. Therefore, upon the mode expansion in terms of the normalizable scalar basis $\phi_{\l,k}$ of both $c$ and $\Sigma'$, the supersymmetric mapping is straightforward.

\section{Summary and discussion}\label{sec:Sumaryanddiscussions}
In this paper, we have  explicitly constructed the map of supersymmetry between the bosonic and fermionic modes of fields in chiral multiplet and vector multiplet on \eads2 by allowing analytic extension of the spectral parameter. For chiral multiplet, the universal shift of the spectrum $\l \rightarrow \l +\i/2$ for spinor field makes the supersymmetry relation manifest. However, for vector multiplet, we need both the $\l + \i/2$ and $\l -\i/2$ in the gaugini spectrum. Additionally, we have found that the point at $\l=0$ of the spectrum remains untouched under the shifting and corresponds to the superpartners of boundary zero modes of vector field.

The presence of the isolated point $\l=0$ in the spectrum  of gaugini is a peculiar aspect of the vector multiplet. 
This spectrum has infinitesimally small measure (finite measure density, but zero measure) in the spectrum and their physical effect seems negligible. This further requires the other bosonic fields such as $\rho$ and $D$ to also have an isolated point in the spectrum, although they may not give additional degrees of freedom. 
However, they do play the role of completing the supersymmetry relation. Especially, boundary zero modes of the vector field have a superpartner due to this spectrum of gaugini. Suppose one consider removing this point in the gaugini spectrum and claim that boundary zero modes of vector field do not have superpartner, then these zero modes would not satisfy the supersymmetry algebra: While acting $Q^2$ on them  would result in zero, the supersymmetry algebra implies a non-zero result as $Q^2 A_{\mu}^{\text{bdry}} =(\cL_v A_{\mu} + \partial_\mu \Sigma) |_{\text{bdry}} = \partial_{\mu}(\sigma +\cosh\eta \,\rho)|_{\text{bdry}} \neq 0$, where $|_{\text{bdry}}$  refers the projection onto the basis function $\partial_\mu\Lambda^{(\ell)}$ given in~\eqref{nonnormalizablescalarsads2}. This is because  the normalizable scalar $\rho$ is multiplied by the exponentially growing function $\cosh\eta$, which certainly includes the non-normalizable function proportional to ~$\Lambda^{(\ell)}$.
Furthermore, in the context of supersymmetric localization \cite{Jeon:2018kec, Iliesiu:2022kny, GonzalezLezcano:2023cuh}, we observe that the superpartner of the boundary zero modes play important role to obtain the correct 1-loop partition functions which is consistent with the perturbative 1-loop result using the standard heat kernel method. 

Up to the aforementioned subtle isolated spectrum, there is an important  implication of our spectral analysis: supersymmetry does not allow normalizable boundary condition for all (bosonic) field in the vector multiplet, while the fields in chiral multiplet can have normalizable boundary condition.
 In this paper, we choose the vector $A_\mu$ and a scalar $\rho$ to be normalizable, but it is unavoidable to allow non-normalizable modes in scalar $\sigma$ and auxiliary field $D$. Therefore, we have used rearrangement of fields into the cohomological form, which can be expanded in terms of  the normalizable basis functions. 
In fact, all cohomological variables can satisfy the normalizable condition. This was indeed the boundary condition assumed in the localization computations performed in \cite{Jeon:2018kec, Iliesiu:2022kny, GonzalezLezcano:2023cuh,Gupta:2015gga,Murthy:2013xpa}, and these computations are now validated by our results. 

Cohomological variables are, by construction, variables with manifest supersymmetry as the elementary fields and their superpartners can be supported by the same basis functions. Thus, we would never be faced with the problem of supersymmetry when dealing with the cohomological variables from the beginning. The question is then, how or to what extent the two formulations in terms of physical variables and cohomological variables describe the same physics. 
We have shed light on the relation between physical and cohomological variables. The normalizable spectrum of fermions in the cohomological variables is clearly obtained from the physical fermions with shifted spectral parameter. The normalizable spectrum of non-elementary bosons is obtained from the non-normalizable scalars $\sigma$ and $D$ through non-trivial field redefinitions  eliminating their non-normalizable part. The isolated spectrums in $\rho$ and $\chi_4$ also need to be removed to obtain the normalizable spectrum in the cohomological variables. 

In order to further test our findings, we could compute the 1-loop contribution to the partition function of the theory at hand using heat kernel technique. This however would require the use of the physical variables which contains some non-normalizable field~$\sigma$. This means that we need an appropriate basis to expand this field, which is necessarily non-normalizable. Adapting the heat kernel technique to include the non-normalizable basis is the subject of current investigation for us and we still do not have a satisfactory way of dealing with it. 
   Therefore, we postpone this issue for the future.

The boundary condition we have imposed (normalizable for all the cohomological variables) may not be the only consistent one.  Since the variational principle in 2-dimensional AdS can in principle allow the non-normalizable modes \cite{Bianchi:2001kw}, we may be able to choose a different set of boundary conditions for each field while still respecting supersymmetry. 
If a physical consideration such as black hole entropy function is given, then the corresponding field configuration and supersymmetric boundary term of the action may determine what boundary behaviors are allowed and thus guide us on what complexified spectrum we should choose. 
In our case the variational principle did not provide us with a good guide. 
Furthermore, if we turn on the interaction between chiral and vector multiplet, then the consistent boundary condition might be modified.

There are other interesting future directions that might be pursued. Firstly, we have not dealt with either the Weyl multiplet or the graviton multiplet, yet, from the perspective of 
black hole physics,
they are the more relevant multiplets to focus on. 
Here, establishing the supersymmetric basis for the gravitons and the gravitini is crucial.
A second direction to explore would be to focus on higher dimensional generalization of our results. Although, in the context of black hole physics, we have highlighted the special role played by \eads2 in the introduction, establishing the supersymmetric basis for higher dimensional AdS is central to the proper definition of general supergravity or string theory. This will extend the utility beyond black hole physics, particularly for AdS$_{d+1}/$CFT$_d$ correspondence. 
\\
\\
\noindent{\bf Acknowledgement:} This work is supported by an appointment to the JRG Program at the APCTP through the Science and Technology 
Promotion Fund and Lottery Fund of the Korean Government and  by the National 
Research Foundation of Korea (NRF) grant funded by the Korea government (MSIT) (No. 2021R1F1A1048531). We wish to thank Sameer Murthy, Leopoldo A. Pando Zayas, Ioannis Papadimitriou, 
and Matthew Roberts  for useful discussions. 


\appendix
\section{Useful definitions} \label{appendixA}
\subsection{Hypergeometric functions}
\paragraph{Definition} We use the following definition for the hypergeometric functions \cite{Gradshteyn:1702455}: 
\be \label{eq:defhypergeo}
F(\alpha\,,\beta\,; \gamma\,; z)= 1 + \frac{\alpha \cdot \beta}{\gamma \cdot 1}z + \frac{\alpha(\alpha +1)\beta(\beta+1)}{\gamma(\gamma+1)\cdot 1\cdot 2}z^2+ \frac{\alpha(\alpha +1)(\alpha +2)\beta(\beta+1)(\beta+2)}{\gamma(\gamma+1)(\gamma+2)\cdot 1\cdot 2 \cdot 3}z^3 +\cdots
\ee
The hypergeometric function also admits the following integral representation : 
\be\label{integralrepF}
F(\alpha\,,\beta\,; \gamma\,; z)=\frac{1}{B(\beta\,, \gamma-\beta)}\int_0^1{\rm d}t~ t^{\beta-1 }(1-t)^{\gamma-\beta-1}(1-tz)^{-\alpha}\,,~~~~[{\rm Re}\, \gamma >{\rm Re} \,\beta>0]\,,
\ee
where
\be
B(\alpha\,,\beta)= \frac{\Gamma(\alpha)\Gamma(\beta)}{\Gamma(\alpha+\beta)}\,.\nn
\ee
\paragraph{Useful relations}~
To explore the asymptotic behavior of fields, we use the following inversion formula :
\begin{eqnarray}\label{Inverse2F1}
&&F(\alpha\,,\beta\,; \gamma\,; z)=\frac{\Gamma(\gamma)\Gamma(\beta-\alpha)}{\Gamma(\beta)\Gamma(\gamma-\alpha)}(-z)^{-\alpha}F\left(\alpha\,,\alpha+1-\gamma\,; \alpha +1 -\beta\,; \frac{1}{z}\right)\nn\\
&&~~~~~~~~~~~~~~~~~~~~~~+\frac{\Gamma(\gamma)\Gamma(\alpha-\beta)}{\Gamma(\alpha)\Gamma(\gamma-\beta)}(-z)^{-\beta}F\left(\beta\,,\beta+1-\gamma\,; \beta +1 -\alpha\,; \frac{1}{z}\right)\,,\\
&&~~~~~~~~~~~~~~~~~~~~~~~~~~~~~~~~~~~~~~~~\textrm{with}~\Big[ |\arg z|< \pi\,,~~\alpha-\beta \neq \pm m\,, ~~m= 0,1,2, \cdots \Big]\,.\nn
\end{eqnarray}
In the actual computations of the integrals in the paper, we have repeatedly made use of Euler's transformation formula, given by: 
\begin{eqnarray}\label{Transformular2}
&&F(\alpha\,,\beta\,; \gamma\,; z)=(1-z)^{\gamma- \alpha-\beta}F(\gamma-\alpha\,, \gamma-\beta\,; \gamma\,; z)\,.
\end{eqnarray}
The following identity is used in the text, which may be derived from Gauss's recursion formulas (see \cite{Gradshteyn:1702455}): 
\begin{align} 
 \g (1+x) F(\a, \b, \g, -x) + x \Big(\a F(\a,+1, \b, \g+1, -x) - \g F(\a, \b, \g, -x)\Big) = \nonumber \\ 
 \frac{ \g}{\a +1 -\b} \Big((1 -\b)F(\a, \b\, \g, -x) +\a  F(\a +1, \b-1, \g, -x) \Big)\, . \label{eq:2f1randomidentity}
\end{align}
 The hypergeometric functions can be integrated as follows: 
\begin{align}\label{integrationF}
\begin{split}
&\int_0^\infty x^{\gamma -1}(x+z)^{-\sigma} F(\alpha\,, \beta\,; \gamma\,; -x)dx \\
&= \frac{\Gamma(\gamma)\Gamma(\alpha - \gamma +\sigma)\Gamma(\beta- \gamma +\sigma)}{\Gamma(\sigma)\Gamma(\alpha +\beta -\gamma +\sigma)} F(\alpha - \gamma +\sigma\,, \beta -\gamma +\sigma\,; \alpha +\beta -\gamma +\sigma \,; 1-z)\,,
\end{split}
\end{align}
provided the following constraints among its parameters are satisfied
\beqa
\Big[ {\rm Re} \,\gamma > 0\,,~~{\rm Re}(\alpha-\gamma +\sigma) > 0\,,~~ {\rm Re}(\beta -\gamma +\sigma)>0\,, ~~|\arg z|< \pi\  \Big]\,.
\eeqa

\subsection{Dirac delta function} The Dirac delta function can be represented using the hypergeometric function. Using the integration formula \eqref{integrationF}, we can express 
\be \label{eq:DiracDeltaIntegralRep}
 \delta(\lambda)= \frac{1}{\pi}\int_0^\infty~dx~  x^{k} F(k+1-\i \lambda \,, k+1 +\i \lambda , k+1,-x)\,.
\ee
This may be proven as follows. In the limit $\epsilon\rightarrow 0$, the above integral may be re-expressed as 
\be
\int_0^\infty  x^k(x+1)^{-\epsilon} F(k+1-\i \lambda \,, k+1 +\i \lambda , k+1\,,-x)= \frac{\Gamma(-\i \lambda+\epsilon)\Gamma(\i\lambda+\epsilon)}{\Gamma(\epsilon)}\,.
\ee

\section{Fermion-Fermion inner product}
\label{appendixB}

In this section of the Appendix we prove that  the inner product $\langle  \psi_{\l^\prime \pm \frac{\i}{2},k^\prime}^{+}\big| \psi_{\l \pm \frac{\i}{2},k}^{+}\rangle$ satisfies Dirac delta-function orthonormality condition.

\subsection{The $\l +\tfrac{\i}{2}$ sector}

Now, from the eigenbasis given in \eqref{EigenAdS2a} and the definition of the inner product in \eqref{innerprodfermion}, after integrating over the angular $\theta$ coordinate and making the change of variable $x = \sinh^2 \tfrac{\eta}{2}$,  we obtain 
\begin{eqnarray}  \label{eq:defalphabeta}
\langle  \psi_{\l^\prime + \frac{\i}{2},k^\prime}^{+}\Big| \psi_{\l + \frac{\i}{2},k}^{+}\rangle & = & A_1 \int \textrm{d} x ~x^k (1+x)^{k+1} F\left(\alpha, \beta, \g, -x\right)F\left( \alpha^\prime, \beta^\prime,\g^\prime, -x\right) + \nonumber \\ 
&& A_2 \int \textrm{d} x ~x^{k+1} (1+x)^{k} F\left(\alpha, \beta, \g+1, -x\right)F\left( \alpha^\prime, \beta^\prime,\g^\prime+1, -x\right) \,, \qquad 
\end{eqnarray} 
where\footnote{We reiterate, here and everywhere in this section, $\l$ itself is real. }, 
\begin{eqnarray} \label{eq:alphabetavalue}
   \a &= &\frac{3}{2} + k - \i \l \, , \quad  \b = \frac{1}{2} + k + \i \l  \,,  \quad  \g = k+1 \nonumber \, ,  \quad  \\ 
    \a^\prime &=& \frac{3}{2} + k - \i \l^\prime \, , \quad \b^\prime = \frac{1}{2} + k + \i \l^\prime \, , \quad \g^\prime = k+1 \, , \\
    A_1 &=& - ~\frac{\i}{(k!)^2} \sqrt{\frac{\Gamma\left(\a\right)\Gamma\left(\b\right) }{\Gamma\left(1-\i \l\right)\Gamma\left(\i \l\right)}} \sqrt{\frac{\Gamma\left(\a^\prime\right)\Gamma\left(\b^\prime\right) }{\Gamma\left(1-\i \l^\prime\right)\Gamma\left(\i \l^\prime\right)}}\, \delta_{k,k^\prime}\, , \nonumber \\
    A_2 &=&  \frac{\i}{(k!)^2} \frac{\left(\frac{1}{2} - \i \l \right)\left(\frac{1}{2} - \i \l^\prime \right)}{\left(1+k\right)^2} \sqrt{\frac{\Gamma\left(\a\right)\Gamma\left(\b\right) }{\Gamma\left(1-\i \l\right)\Gamma\left(\i \l\right)}} \sqrt{\frac{\Gamma\left(\a^\prime\right)\Gamma\left(\b^\prime\right) }{\Gamma\left(1-\i \l^\prime\right)\Gamma\left(\i \l^\prime\right)}}\, \delta_{k,k^\prime}\,  \nonumber \, \\
    &=& -~ \frac{\left(\gamma - \beta\right)\left(\gamma^\prime - \b^\prime\right)}{\gamma \gamma^\prime} A_{1} \, .
    \label{eq:identityA1A2}
\end{eqnarray} 
\subsection*{The first integral}
Let us introduce
\begin{align}
    \cI_1 \equiv A_1 \int \textrm{d} x ~x^k (1+x)^{k+1} F\left(\alpha, \beta, \g, -x\right)F\left( \alpha^\prime, \beta^\prime,\g^\prime, -x\right)\, .
\end{align}
Using \eqref{Transformular2} we have:
\begin{align}
    \cI_1 = A_1 \int \textrm{d} x ~ x^k F\left(\g - \a, \g - \b, \g, -x\right)F\left( \alpha^\prime, \beta^\prime,\g^\prime, -x\right)\, .
\end{align}
Now let us use the integral representation for the second hypergeometric function to obtain 
\begin{align}
      \cI_1 = \frac{A_1}{B\left(\b^\prime , \g^\prime - \b^\prime\right)} \int \textrm{d} x ~ \int_0^1 \textrm{d} t  ~ t^{\b^\prime - 1} \left(1-t\right)^{\g^\prime - \b^\prime -1} t^{- \a^\prime} \left(\frac{1}{t} + x\right)^{- \a^\prime} x^k F\left(\g - \a, \g - \b, \g, -x\right)
\end{align}
for which we only need $ \text{Re}( \g^\prime) > \text{Re}(\b^\prime) >0$ which is always satisfied for $k >0$. We perform the $x$ integral using \eqref{integrationF}:
\begin{align}
        \cI_1 = \frac{A_1}{B\left(\b^\prime , \g^\prime - \b^\prime\right)} \int \textrm{d} x ~ \int_0^1 \textrm{d} t  ~ t^{\b^\prime - 1} \left(1-t\right)^{\g^\prime - \b^\prime -1} t^{- \a^\prime} \left(\frac{1}{t} + x\right)^{- \a^\prime} x^k F\left(\g - \a, \g - \b, \g, -x\right)
\end{align}
for which we need:
\begin{align}
    \text{Re}(\g) >0\,, \quad  \text{Re}(\g - \a - \g + \a^\prime ) = \text{Re}(\a^\prime - \a) >0\, \quad \text{Re}(\g - \b - \g + \a^\prime ) = \text{Re}(\a^\prime - \b) >0\, .
\end{align}
They are all satisfied except for the second one that is equal to zero.  We then propose the regularization $\alpha^\prime \rightarrow \alpha^\prime + \epsilon$ with $\epsilon >0$. To satisfy $\gamma-\alpha-\beta = \gamma^\prime - \alpha^\prime - \beta^\prime$,  we also deform $\b^\prime \rightarrow \b^\prime + \epsilon $ and $\g^\prime \rightarrow \g^\prime  + 2 \epsilon$. Therefore, our regularization scheme includes the following re-definitions
\begin{eqnarray}
    \a \rightarrow \a\,, \, \,  \b \rightarrow \b\, , \,\,  \g \rightarrow \g\, \,\,; \quad \a^\prime \rightarrow \a^\prime + \epsilon\, , \,\, \b^\prime \rightarrow \b^\prime + \epsilon\, , \,\,\g^\prime \rightarrow \g^\prime + 2\epsilon\, .
\end{eqnarray} We then have:
\begin{align}
    \cI_1 & = \frac{A_1}{B\left(\b^\prime +\epsilon , \g^\prime - \b^\prime +\epsilon 
\right)}\int_0^1 \textrm{d} t  ~ t^{\b^\prime - \a^\prime - 1} \left(1-t\right)^{\g^\prime +\epsilon - \b^\prime -1} \frac{\Gamma(\g) \Gamma(\a^\prime +\epsilon - \a)\Gamma(\a^\prime + \epsilon- \beta)}{\Gamma(\a^\prime +\epsilon ) \Gamma(\a^\prime +\epsilon +\g - \a -\b)}\nonumber  \\
    & \times F\left(\a^\prime +\epsilon - \a;\a^\prime+\epsilon - \beta ,\a^\prime +\epsilon +\g - \a - \b   ; 1- \frac{1}{t}\right)\, .
\end{align}
Implementing the change of variables: $x = \frac{1}{t}- 1$,  we have:
\begin{align}
    \cI_1 & = \frac{A_1}{B\left(\b^\prime +\epsilon , \g^\prime +\epsilon - \b^\prime\right)} \int_0^\infty\textrm{d} x ~ x^{\g^\prime - \b^\prime -1 +\epsilon} (1+x)^{\a^\prime -\epsilon - \g^\prime}\frac{\Gamma(\g) \Gamma(\a^\prime +\epsilon - \a)\Gamma(\a^\prime + \epsilon- \beta)}{\Gamma(\a^\prime +\epsilon ) \Gamma(\a^\prime +\epsilon +\g - \a -\b)} \nonumber \\
    & \times F\left(\a^\prime +\epsilon - \a;\a^\prime +\epsilon - \beta ,\a^\prime +\epsilon + \g- \a - \b   ; -x\right) \, .
\end{align}
By virtue of the relation $\g - \a - \b = \g^\prime - \a^\prime - \b^\prime$, which we have been careful not to violate even after the regularization process introduced above, we can write:
\begin{align}
      \cI_1 & =  \frac{A_1 \Gamma(\g^\prime +2\epsilon )}{ \Gamma(\b^\prime +\epsilon) \Gamma(\g^\prime +\epsilon  - \b^\prime)}  \frac{\Gamma(\g) \Gamma(\a^\prime +\epsilon - \a)\Gamma(\a^\prime + \epsilon - \beta)}{\Gamma(\a^\prime +\epsilon ) \Gamma(\g^\prime +\epsilon -\b^\prime)} \nonumber \\ & \times 
      \int_0^\infty\textrm{d} x ~ x^{\g^\prime +\epsilon- \b^\prime -1} (1+x)^{\a^\prime -\epsilon- \g^\prime} F\left(\a^\prime +\epsilon - \a;\a^\prime +\epsilon - \beta ,\g^\prime +\epsilon -\b^\prime   ; -x\right) \, .
\end{align}
 Now using \eqref{integrationF} we have:
\begin{align}
    \cI_1 = \frac{A_1}{B\left(\b^\prime , \g^\prime - \b^\prime\right)} \frac{\Gamma(\g) \Gamma(\a^\prime +\epsilon - \a)\Gamma(\a^\prime + \epsilon - \beta)}{\Gamma(\a^\prime +\epsilon ) } \frac{ \Gamma(\b^\prime +\epsilon - \a)\Gamma(\b^\prime + \epsilon - \beta)}{\Gamma(\g^\prime -\a^\prime ) \Gamma(\a^\prime +\b^\prime +2\epsilon - \a - \b )}\, .
\end{align} For the convergence of the above integral, we need 
\beqa \label{eq:randomeqn}
\text{Re}(\g^\prime - \b^\prime) > 0\, \quad \text{Re}(\b^\prime + \epsilon - \a) > 0\,, \quad \text{Re}(\b^\prime +\epsilon - \beta) >0
\eeqa but the middle one implies $\epsilon> 1
$ as can be checked from \eqref{eq:defalphabeta}. But we will see from the next half of the computation that the divergence (which this finite $\epsilon$ seemingly regulates) nicely cancels in the final sum and therefore, to compute the actual inner product we do not need any finite valued regulator. This is seen in the next computation and given its importance commented again after equation \eqref{eq:constraints}. 

\subsection*{The second integral}

Let us now consider $\cI_2$ defined as:
\begin{align}
    \cI_2 \equiv A_2 \int \textrm{d} x ~x^{k+1} (1+x)^{k} F\left(\alpha, \beta, \g +1, -x\right)F\left( \alpha^\prime, \beta^\prime,\g^\prime +1 , -x\right)\,.
\end{align}
Using \eqref{Transformular2} we have:
\begin{align}
    \cI_2 = A_2 \int \textrm{d} x ~ x^{k+1} F\left(\g+1 - \a, \g+1 - \b, \g+1, -x\right)F\left( \alpha^\prime, \beta^\prime,\g^\prime +1, -x\right)\,.
\end{align}
Now let us use the integral representation for the second hypergeometric function. We obtain 
\begin{align}
\begin{split}
      \cI_2 & = \frac{A_2}{B\left(\b^\prime , \g^\prime+1 - \b^\prime\right)} \int \textrm{d} x ~ \int_0^1 \textrm{d} t  ~ t^{\b^\prime - 1} \left(1-t\right)^{\g^\prime - \b^\prime } t^{- \a^\prime} \left(\frac{1}{t} + x\right)^{- \a^\prime} \\ 
      &  \times x^{k+1}  F\left(\g +1 - \a, \g +1  - \b, \g +1, -x\right)\,
      \end{split}
\end{align}
for which we only need $ \text{Re}( \g^\prime) > \text{Re}(\b^\prime) >0$ which is always satisfied for $k >0$. Now let us perform the $x$ integral using \eqref{integrationF}:
\begin{align}
\begin{split}
        \cI_2 &  = \frac{A_2}{B\left(\b^\prime , \g^\prime+1 - \b^\prime\right)} \int \textrm{d} x ~ \int_0^1 \textrm{d} t  ~ t^{\b^\prime - 1} \left(1-t\right)^{\g^\prime - \b^\prime } t^{- \a^\prime} \left(\frac{1}{t} + x\right)^{- \a^\prime} \\
        & \times  x^{k+1} F\left(\g+1 - \a, \g+1  - \b, \g +1, -x\right)
        \end{split}
\end{align}
for which we need:
\begin{align}
    \text{Re}(\g) >0\,, \quad  \text{Re}(\g - \a - \g + \a^\prime ) = \text{Re}(\a^\prime - \a) >0\, , \quad \text{Re}(\g - \b - \g + \a^\prime ) = \text{Re}(\a^\prime - \b) >0\, ,
\end{align}
which are all satisfied except for the second one which is now equal to zero. We perform the same regularization as for $\mathcal{I}_1$ above by introducing a regulator $\epsilon$ and obtain
\begin{align}
\begin{split}
    \cI_2 & = \frac{A_2}{B\left(\b^\prime +\epsilon, \g^\prime +\epsilon+1 - \b^\prime\right)}\int_0^1 \textrm{d} t  ~ t^{\b^\prime  - \a^\prime - 1} \left(1-t\right)^{\g^\prime +\epsilon  - \b^\prime } \frac{\Gamma(\g +1) \Gamma(\a^\prime +\epsilon - \a)\Gamma(\a^\prime + \epsilon- \beta)}{\Gamma(\a^\prime +\epsilon ) \Gamma(\a^\prime +\epsilon +\g +1  - \a -\b)} \\
    & \times F\left(\a^\prime +\epsilon - \a;\a^\prime+\epsilon - \beta ,\a^\prime +\epsilon +\g +1  - \a - \b   ; 1- \frac{1}{t}\right)\, . 
    \end{split}
\end{align}
On changing variables: $x = \frac{1}{t}- 1$, we have
\begin{align}
\begin{split}
   & \cI_2  = \frac{A_2}{B\left(\b^\prime +\epsilon , \g^\prime +\epsilon - \b^\prime\right)} \int_0^\infty\textrm{d} x ~ x^{\g^\prime +\epsilon - \b^\prime } (1+x)^{\a^\prime - \epsilon - \g^\prime -1} ~ \times\\
    &  \frac{\Gamma(\g +1 ) \Gamma(\a^\prime +\epsilon - \a)\Gamma(\a^\prime + \epsilon- \beta)}{\Gamma(\a^\prime +\epsilon ) \Gamma(\a^\prime +\epsilon +\g +1 - \a -\b)}  F\left(\a^\prime +\epsilon - \a;\a^\prime +\epsilon - \beta ,\a^\prime +\epsilon + \g+1- \a - \b   ; -x\right) \, .
    \end{split}
\end{align}
Again, since $\g - \a - \b = \g^\prime - \a^\prime - \b^\prime$ we can write
\begin{align}
\begin{split}
     & \cI_2  = \frac{A_2}{B\left(\b^\prime +\epsilon , \g^\prime +\epsilon +1 - \b^\prime\right)} \int_0^\infty\textrm{d} x ~ x^{\g^\prime +\epsilon - \b^\prime } (1+x)^{\a^\prime - \epsilon - \g^\prime-1} ~ \times \\
      &  \frac{\Gamma(\g +1 ) \Gamma(\a^\prime +\epsilon - \a)\Gamma(\a^\prime + \epsilon - \beta)}{\Gamma(\a^\prime +\epsilon ) \Gamma(\g^\prime +\epsilon +1  -\b^\prime)}  F\left(\a^\prime +\epsilon - \a;\a^\prime +\epsilon - \beta ,\g^\prime +\epsilon -\b^\prime +1  ; -x\right) \, \\
      & \times  \int_0^\infty\textrm{d} x ~ x^{\g^\prime - \b^\prime } (1+x)^{\a^\prime - \g^\prime-1} F\left(\a^\prime +\epsilon - \a;\a^\prime +\epsilon - \beta ,\g^\prime -\b^\prime +1  ; -x\right) \,  \\
      &  =   \frac{A_2 \Gamma(\g^\prime +2 \epsilon +1 )}{ \Gamma(\b^\prime +\epsilon) \Gamma(\g^\prime +\epsilon +1 - \b^\prime)}  \frac{\g}{ \g^\prime +\epsilon - \b^\prime}
      \frac{\Gamma(\g ) \Gamma(\a^\prime +\epsilon - \a)\Gamma(\a^\prime + \epsilon - \beta)}{\Gamma(\a^\prime +\epsilon ) \Gamma(\g^\prime  +\epsilon  -\b^\prime)}  
      \\& \times  \int_0^\infty\textrm{d} x ~ x^{\g^\prime +\epsilon - \b^\prime } (1+x)^{\a^\prime - \epsilon- \g^\prime-1} F\left(\a^\prime +\epsilon - \a;\a^\prime +\epsilon - \beta ,\g^\prime +\epsilon -\b^\prime +1  ; -x\right)\,.
      \end{split}
      \end{align}
Now using \eqref{integrationF} we have
\begin{align}
    \cI_2 = \frac{A_2}{B\left(\b^\prime , \g^\prime +1  - \b^\prime\right)} \frac{\Gamma(\g +1) \Gamma(\a^\prime +\epsilon - \a)\Gamma(\a^\prime + \epsilon - \beta)}{\Gamma(\a^\prime +\epsilon ) \Gamma(\g^\prime - \b^\prime +1 )} \frac{ \Gamma(\b^\prime +\epsilon - \a)\Gamma(\b^\prime + \epsilon - \beta)}{ \Gamma(\a^\prime +\b^\prime +2\epsilon - \a - \b )}\, ,
\end{align}
 where for the convergence of the integral, we need $\text{Re}(\g^\prime - \b^\prime) > 0\, \quad \text{Re}(\b^\prime + \epsilon - \a) > 0\,, \quad \text{Re}(\b^\prime +\epsilon - \beta) >0.$ Note that, here too, the middle inequality  requires $\epsilon > 1$.

Using the relation between $A_1$ and $A_2$ given in \eqref{eq:identityA1A2}, we can combine the two integrals and get 
\begin{align}
    &\cI_1 +\cI_2  =  A_1 \frac{ \g (\g^\prime+2 \epsilon) \Gamma(\g^\prime +2 \epsilon  )}{ \Gamma(\b^\prime  +\epsilon) \Gamma(\g^\prime  +\epsilon  - \b^\prime)}  \frac{\Gamma(\g) \Gamma(\a^\prime +\epsilon - \a)\Gamma(\a^\prime + \epsilon - \beta)}{\Gamma(\a^\prime +\epsilon ) \Gamma(\g^\prime +\epsilon -\b^\prime)} ~ \times \nonumber \\ & \Biggl[
      \int_0^\infty\textrm{d} x ~ x^{\g^\prime  +\epsilon - \b^\prime -1} (1+x)^{\a^\prime - \epsilon - \g^\prime - 1} \Biggl( (\g^\prime +\epsilon - \b^\prime) (1+x) F\left(\a^\prime +\epsilon - \a;\a^\prime +\epsilon - \beta ,\g^\prime  +\epsilon -\b^\prime   ; -x\right) \nonumber \\
      & -   (\g- \b)  x F\left(\a^\prime +\epsilon - \a,\a^\prime +\epsilon - \beta ,\g^\prime  +\epsilon -\b^\prime +1  ; -x\right)\Biggr)\Biggr]\, . 
\end{align} Then \eqref{eq:2f1randomidentity} gives us
\begin{align} \label{eq:sumI1I2}
    \cI_1 +\cI_2  &=   A_1 \frac{ \g (\g^\prime+2 \epsilon) \Gamma(\g^\prime +2 \epsilon  )}{ \Gamma(\b^\prime  +\epsilon) \Gamma(\g^\prime  +\epsilon  - \b^\prime)}  \frac{\Gamma(\g) \Gamma(\a^\prime +\epsilon - \a)\Gamma(\a^\prime + \epsilon - \beta)}{\Gamma(\a^\prime +\epsilon ) \Gamma(\g^\prime +\epsilon -\b^\prime)} \times\nonumber \\
    &  \Biggl[
      \int_0^\infty\textrm{d} x ~ x^{\g^\prime +\epsilon- \b^\prime -1} (1+x)^{\a^\prime - \g^\prime - \epsilon - 1} \Biggl( \frac{(\g^\prime +\epsilon - \b^\prime)}{\b -\a +1} \Bigl((\b+1- \a^\prime -\epsilon) \times \nonumber \\
      & F\left(\a^\prime +\epsilon - \a;\a^\prime +\epsilon - \beta ,\g^\prime +\epsilon -\b^\prime   ; -x\right)   \\
      & +   (\a^\prime +\epsilon - \a)   F\left(\a^\prime+1 +\epsilon - \a,\a^\prime +\epsilon - \beta-1 ,\g^\prime +\epsilon -\b^\prime  ; -x\right)\Bigr)\Biggr)\Biggr]\,.  \nonumber
\end{align} The above integrals are in the standard integral form, as described in equation \eqref{integrationF}, subject to the constraints as described there. We first check the constraints. They are 
\begin{subequations} \label{eq:constraints}
    \begin{align}
    \text{Re}(\g^\prime -\b^\prime) & > 0\,, \\
    \text{Re}(\a^\prime +\epsilon - \a - \g^\prime + \b^\prime +1+\g^\prime - \a^\prime) & = \text{Re}(\b^\prime - \a +1 +\epsilon) =\epsilon >0. \\
  \text{Re}(\a^\prime + \epsilon - \beta - \g^\prime + \b^\prime +1 +\g^\prime- \a^\prime) & = \text{Re}(\b^\prime - \b+1) = \epsilon +1 >0 \\
\text{Re}(\a^\prime +1 + \epsilon - \a - \g^\prime +\b^\prime +1 +\g^\prime -\a^\prime)  & = \text{Re}(\b^\prime - \a +2 +\epsilon) =2 +\epsilon >0 \\
\text{Re}(\a^\prime + \epsilon - \b - 1 -\g^\prime +\b^\prime +1 + \g^\prime - \a^\prime)& = \text{Re}(\b^\prime - \b +\epsilon) = \epsilon >0\, .
\end{align}
\end{subequations}
Using \eqref{eq:alphabetavalue}, we see that all of the above are satisfied as long as $\epsilon >0$. This point is important and needs a reiteration. As remarked after \eqref{eq:randomeqn}, the sum of $\cI_1$ and $\cI_2$ does \textit{not} need any finite regulator. This is because in the sum the divergent pieces nicely cancel each other as can be checked using identity \eqref{eq:2f1randomidentity}. Therefore, we can now consistently take  the limit $\epsilon\rightarrow 0$.

The upshot of the calculation is that we can write the inner product as 
\begin{align}
    \cI_1+\cI_2 & =  A_1  \frac{\g (\g^\prime+2 \epsilon) \Gamma(\g^\prime +2 \epsilon  )}{ \Gamma(\b^\prime  +\epsilon) \Gamma(\g^\prime  +\epsilon  - \b^\prime)}  \frac{\Gamma(\g) \Gamma(\a^\prime +\epsilon - \a)\Gamma(\a^\prime + \epsilon - \beta)}{\Gamma(\a^\prime +\epsilon ) \Gamma(\g^\prime +\epsilon -\b^\prime)}   \nonumber \\ 
    & \times \Biggl[\frac{(\g^\prime +\epsilon - \b^\prime)}{\b -\a +1} \Bigl((\b+1- \a^\prime -\epsilon) \nonumber \\
      & \times\frac{\Gamma(\g^\prime +\epsilon - \b^\prime)\Gamma(\b^\prime - \a +1 +\epsilon)\Gamma(\b^\prime - \b + 1 +\epsilon)}{\Gamma(1 + \g^\prime +\epsilon -\a^\prime)\Gamma(\a^\prime + \b^\prime - \a +1 +2 \epsilon)} \nonumber \\
      & +   (\a^\prime +\epsilon - \a)   \frac{\Gamma(\g^\prime +\epsilon - \b^\prime)\Gamma(\b^\prime - \a +2 +\epsilon)\Gamma(\b^\prime - \b  +\epsilon)}{\Gamma(1 + \g^\prime +\epsilon -\a^\prime)\Gamma(\a^\prime + \b^\prime - \a +1 +2 \epsilon)}\Bigr)\Biggr)\Biggr]\,
\end{align} without needing any \textit{finite} regulator.

So far we have been schematic. We now plugging the explicit values of $\{\alpha, \alpha^\prime,\beta, \beta^\prime ,\g,\g^\prime\}$ as given in \eqref{eq:alphabetavalue} and take $\epsilon \rightarrow 0$ limit. We see that the singular pieces combine nicely together to give us a delta function. This uses the representation of delta function as given in  \eqref{eq:DiracDeltaIntegralRep} and we have
\begin{align} \label{eq:ffinnerprod_sameshift}
    \langle  \psi_{\l^\prime + \frac{\i}{2},k^\prime}^{+}\Big| \psi_{\l + \frac{\i}{2},k}^{+}\rangle = \delta_{k,k^\prime} \delta(\lambda - \lambda^\prime)\, . 
\end{align}
\subsection{The $\l - \tfrac{\i}{2}$ sector}

Now, as discussed in the beginning of this section, we show that the separate calculation for the other combinations of shifts is unnecessary and may directly be extracted from this result in \eqref{eq:ffinnerprod_sameshift}. This follows from the observation that
\begin{eqnarray} \label{eq:Identity_othershifts} \langle  \psi_{\l^\prime - \frac{\i}{2},k^\prime}^{+}\big| \psi_{\l - \frac{\i}{2},k}^{+}\rangle = \langle  \psi_{\l^\prime + \frac{\i}{2},k^\prime}^{+}\big| \psi_{\l + \frac{\i}{2},k}^{+}\rangle \Bigg|_{\{\l \rightarrow -\l, \l^\prime \rightarrow -\l^\prime\}}\,,
\end{eqnarray} 
and 
\begin{subequations} \label{eq:oppositeshifting}
\begin{eqnarray}  \langle  \psi_{\l^\prime + \frac{\i}{2},k^\prime}^{+}\big|\psi_{\l - \frac{\i}{2},k}^{+}\rangle = \langle  \psi_{\l^\prime + \frac{\i}{2},k^\prime}^{+}| \psi_{\l + \frac{\i}{2},k}^{+}\rangle \Bigg|_{\{\l \rightarrow -\l, A_2 \rightarrow - A_2\}} \sim  \,\, (\l+\l^\prime) \delta_{k,k^\prime} \delta(\l+ \l^\prime)\,,\qquad  
\end{eqnarray}
\begin{eqnarray}  
\langle  \psi_{\l^\prime - \frac{\i}{2},k^\prime}^{+}\big| \psi_{\l + \frac{\i}{2},k}^{+}\rangle = \langle  \psi_{\l^\prime + \frac{\i}{2}k^\prime}^{+}\big| \psi_{\l + \frac{\i}{2},k}^{+}\rangle \Bigg|_{\{\l^\prime \rightarrow -\l^\prime, A_2 \rightarrow - A_2\}} \sim   \,\, (\l+\l^\prime) \delta_{k,k^\prime}\delta(\l+ \l^\prime) \,,\qquad  
\end{eqnarray}
\end{subequations}  where the $\sim$ symbol denotes the presence of some overall constant pre-factors that are finite and do not alter the behavior of these inner products. Note that, unlike before, the calculations in \eqref{eq:oppositeshifting} do indeed require a finite regulator.  With this caveat, the identities above allow us to proceed exactly in the same manner to what we already have and we obtain 
\begin{subequations} \label{eq:Results_othershifts}
\begin{eqnarray}  \langle  \psi_{\l^\prime - \frac{\i}{2},k^\prime}^{+}\big| \psi_{\l - \frac{\i}{2},k}^{+}\rangle = \delta_{k,k^\prime} \delta(\lambda - \lambda^\prime)\,,
\end{eqnarray} 
\begin{eqnarray}  \langle  \psi_{\l^\prime + \frac{\i}{2},k^\prime}^{+}\big|\psi_{\l - \frac{\i}{2},k}^{+}\rangle \,\, \sim \,\, (\l+\l^\prime) \delta_{k,k^\prime} \delta(\l+ \l^\prime) = 0 \, , 
\end{eqnarray}
\begin{eqnarray}  \langle  \psi_{\l^\prime - \frac{\i}{2},k^\prime}^{+}\big| \psi_{\l + \frac{\i}{2},k}^{+}\rangle \,\, \sim  \,\, (\l+\l^\prime) \delta_{k,k^\prime}\delta(\l+ \l^\prime) = 0 \, . 
\end{eqnarray}
\end{subequations}
For the other pair of eigenmodes, we simply note that 
\begin{eqnarray}
    \langle  \psi_{\l^\prime,k^\prime}^{-}\big| \psi_{\l ,k}^{-}\rangle = - \i \left(\psi_{\l^\prime,k^\prime}^{+}\right)^T C \psi_{\l ,k}^{-} = \i \left(\psi_{\l,k}^{-}\right)^T C \psi_{\l^\prime ,k^\prime}^{+} = \langle  \psi_{\l,k}^{+}\big| \psi_{\l^\prime ,k^\prime}^{+}\rangle
\end{eqnarray} which implies
\begin{subequations} \label{eq:Results_otherothershifts}
\begin{eqnarray}  \langle  \psi_{\l^\prime - \frac{\i}{2},k^\prime}^{-}\big| \psi_{\l - \frac{\i}{2},k}^{-}\rangle = \delta_{k,k^\prime} \delta(\lambda - \lambda^\prime)\,,
\end{eqnarray} 
\begin{eqnarray}  \langle  \psi_{\l^\prime + \frac{\i}{2},k^\prime}^{-}\big|\psi_{\l - \frac{\i}{2},k}^{-}\rangle  \,\, \sim \,\, (\l+\l^\prime) \delta_{k,k^\prime} \delta(\l+ \l^\prime) = 0 \, , 
\end{eqnarray}
\begin{eqnarray}  \langle  \psi_{\l^\prime - \frac{\i}{2},k^\prime}^{-}\big| \psi_{\l + \frac{\i}{2},k}^{-}\rangle \,\, \sim  \,\, (\l+\l^\prime) \delta_{k,k^\prime}\delta(\l+ \l^\prime) = 0 \, . 
\end{eqnarray}
\end{subequations}
This concludes our computation of the inner products between every combinations of the shifted fermions basis. 

   \section{Supersymmetry transformation}
   \subsection{Physical variables}
We consider a gauge neutral chiral multiplet consisting of two complex scalars, two Dirac spinors and two auxiliary fields $\{\phi\,,\overline{\phi}\,, \psi\,,\overline{\psi}\,, F\,, \overline{F}\}$, with $R$-charge $(r,-r, r-1, -r+1, r-2, -r+2)$.  There supersymmetric variation is given as
\begin{align} \label{eq:Qchiral01}
\begin{split}
    Q \phi & = \bar\epsilon \psi\,\\
Q \bar\phi & = \epsilon \bar\psi  \\
Q \psi & = \i \gamma^\mu \epsilon \partial_\mu \phi  
+\i \frac{\r}{2} \epsilon\phi   +\bar\epsilon F
 \\
Q \bar\psi & = \i \gamma^\mu \bar\epsilon \partial_\mu \bar\phi  
+\i \frac{\r}{2} \bar\epsilon  \bar\phi    +\epsilon \bar F 
\\
Q F & =    \i \epsilon\gamma^\mu D_\mu \psi  -\i\frac{\r}{2}  \epsilon\psi 
\\
Q \bar F & =    \i \bar\epsilon\gamma^\mu D_\mu \bar\psi  -\i\frac{\r}{2}\bar\epsilon\bar\psi \,.
\end{split}
\end{align}

We also consider an Abelian vector multiplet consists of a vector, two real scalars, two Dirac spinors and an auxiliary scalar, $\{A_\mu\,,\sigma\,,\rho\,,\chi\,,\overline{\chi}\,,D \}$, whose supersymmetric variation is given by 
%
   \beqa
\label{eq:deltaA}
Q A_\mu &=& - \frac{\i}{2} (\bar\epsilon \gamma_\mu \chi + \epsilon \gamma_\mu \bar\chi) 
\nn 
\\
Q \S  &=& -\frac{1}{2}  (\bar\epsilon \chi - \epsilon  \bar\chi) \nn\\
Q \P  &=& - \frac{\i}{2}  (\bar\epsilon \gamma_3 \chi + \epsilon \gamma_3 \bar\chi)\\
Q \chi 
&=& \i \gamma_3 \epsilon F_{12} -{D} \epsilon - \i \gamma^\mu  \epsilon D_\mu\S  - \gamma_3 \gamma^\mu D_\mu (\epsilon \P  )\nn
\\
Q \bar\chi
 &=& \i \gamma_3 \bar\epsilon F_{12} +{D} \bar\epsilon +
 \i \gamma^\mu  \bar\epsilon  D_\mu\S  - \gamma_3 \gamma^\mu D_\mu ( \bar\epsilon  \P  )\nn
 \\
Q {D}
&=& - \frac{\i}{2}   \left(\bar\epsilon \gamma^\mu D_\mu\chi - \frac{\i}{2}  \epsilon \gamma^\mu  D_\mu\bar\chi \right) \nn
\,,
\eeqa
where $F_{12}= \frac{1}{2} \epsilon^{\mu\nu}F_{\mu\nu}$. The supersymmetry algebra is closed to the Lie derivative and the gauge transformation as 
\be
Q^2 \= \i \cL_v + \i \delta_G (\Sigma)\,,
\ee
where the Killing vector $v^\mu$ and the field dependent gauge parameter  are given by 
\be
v^\mu = \bar{\epsilon}\gamma^\mu \epsilon\,,\quad \Sigma \= -v^\mu A_\mu +\i \bar{\epsilon}\epsilon \sigma -  \bar\epsilon \gamma_3 \epsilon \rho\,.
\ee
In \eads2, with the Killing spinors in \eqref{eq:KillingSpinors}, these are given by 
\be\label{LambdaParameter}
v^\mu \partial_\mu = \partial_\theta\,,\qquad 
\Sigma \= -A_\theta - \sigma - \i \cosh\eta\, \rho\,.
\ee
One can show that $Q \Sigma =0$ identically using Fierz identity.

For the systematic treatment of gauge fixing process, we include the ghost multiplet consisting of ghost, anti-ghost and auxiliary field $\{ c\,,\bar{c}\,,b\}$, with the only non-trivial BRST transformation being
\be
Q_{\text{brst}} A_\mu = \partial_\mu c\,,~\qquad Q_{\text{brst}} \bar{c}= b\,.
\ee
To encode the ghost multiplet into the supermultiplet, we impose the supersymmetry variation is given by
\be
Q c = -\Sigma +\Sigma_0\,,\qquad Q b = \i v^\mu \partial_\mu \bar{c}
\ee
where $\Sigma_0$ is non-normalizable part of $\Sigma$ because the ghost field $c$ is normalizable scalar as the gauge parameter is by definition normalizable scalar.
Then, the vector together with the ghost multiplet form a supermultiplet not by the supersymmetry $Q$ but by the following combined charge
\be
\qeq \equiv Q + Q_{\text{brst}}
\ee
which we call equivariant supercharge because its algebra equivariantly closes as
\be
\qeq^2 \= \i \cL_v + \i \delta_G (\Sigma_0)\,.
\ee

\subsection{Cohomological variables} \label{app:twistedvariable}
It is convenient to reorganize the fields into a certain representation of supersymmetry called ``cohomological variables''. The cohomological variables consist of $\qeq$-cohomology complex $(\Phi, \qeq\Phi, \Psi,\qeq\Psi)$, where we call $\Phi$ and $\Psi$ the elementary boson and fermion, and $\qeq\Phi$ and $\qeq\Psi$ are their superpartners. 

For chiral multiplet, we define twisted variables for the spinor fields $\psi$ and $\overline\psi$ as
\be
\epsilon \psi\,,\quad \epsilon \overline{\psi}\,,\quad \bar\epsilon \psi\,,\quad \bar{\epsilon}\overline{\psi}\,.
\ee
The inverse relation is 
\be
\psi = \frac{1}{\bar{\epsilon}\epsilon}\Bigl( \epsilon (\bar\epsilon \psi) - \bar{\epsilon}(\epsilon \psi)\Bigr)
\,, \qquad \overline{\psi} = \frac{1}{\bar{\epsilon}\epsilon}\Bigl(  \epsilon (\bar{\epsilon}\overline{\psi} - \bar{\epsilon}(\epsilon\overline{\psi})) \Bigr)\,.
\ee
Then the cohomogical variables are 
\begin{equation}
\begin{split}
&\Phi  =  \{\phi\, ,\bar\phi\}\, , \qquad\qquad \qeq \Phi =\{ \qeq \phi \, ,\qeq \bar{\phi}\}\,  \label{eq:CQarray} ,\\ 
&\Psi  =  \{\epsilon \psi \, , \bar{\epsilon} \bar{\psi}\}\,, \qquad\quad\qeq \Psi = \{\qeq \left(\eps \psi\right)\, , \qeq \left( \bar{\epsilon} \bar{\psi}\right)\}\, .    
\end{split}
\end{equation}
The explicit expression of the variables in $\qeq \Phi$ and $\qeq \Psi$ are
\begin{align}\label{eq:ChBPS3}
\begin{split}
\qeq\phi& = \bar{\epsilon} \psi \,, 
\\
\qeq \bar{\phi} & = \epsilon \bar{\psi},
\\
\qeq(\epsilon \psi) & = \epsilon \bar{\epsilon} F + \i \epsilon \gamma^{\mu} \epsilon \partial_{\mu} \phi, \\
 \qeq(\bar{\epsilon} \bar{\psi}) & = \bar{\epsilon}\epsilon \bar{F} + \i \bar{\epsilon} \gamma^{\mu}  \bar{\epsilon} \partial_{\mu} \bar{\phi} .\\
 \end{split}
 \end{align}
 For a neutral chiral multiplet $\qeq = Q$.
 
For the vector multiplet, with $\gamma_A =(\gamma_1\,,\gamma_2\,,\gamma_3\,,1) $, we define twisted variables for the gaugini $\chi$ and $\overline{\chi}$ as  
\be  \label{eq:ChiA}
\chi_A = -\frac{\i}{2} (\overline{\epsilon}\gamma_A \chi + \epsilon \gamma_A \overline{\chi})\,,\quad A= 1,2,3,4\,.
\ee
The inverse relation is
\be
\chi = -\frac{1}{\i\bar{\epsilon}\epsilon} \gamma^A \epsilon \chi_A\,,\qquad \overline{\chi}= \frac{1}{\i\bar{\epsilon}\epsilon}\gamma^A \bar{\epsilon} \chi_A\,.
\ee
Then the cohomological variables are
\begin{equation}
\begin{split}
&    \Phi  =  \{A_{a}\,,  \P\,\}\,, \hspace{7mm} \qquad \qeq \Phi =\{ \qeq A_a \, ,  \qeq \P\,\}\,, \label{eq:VQarray} \\
&\Psi  =  \{\chi_4\,, c\,, \bar{c}\}\,,\hspace{5mm}\qquad \qeq \Psi = \{\qeq \chi_4\,, \qeq c \, ,\qeq \bar{c}\} ,\quad \quad \left\{\Sigma_0 \right\}\,,
\end{split}
\end{equation}
where we note that $\Sigma_0$ is a singlet in the cohomological representation. 
The explicit expression of the variables in $\qeq \Phi$ and $\qeq \Psi$ are given as
   \begin{eqnarray}
\qeq A_{\mu}&=&
\chi_{\mu} + \partial_{\mu} c \,, \label{eq:QLamb4} \nn\\
\qeq \P  & = &
\chi_3\,, \nn \\
\qeq \chi_4 &=&\i \bar{\epsilon}\epsilon D+ \i  \bar\epsilon \gamma_3 \epsilon  \left(-\i F_{12} 
+  \P  \right) + \i \bar\epsilon \gamma_3 \gamma^\mu \epsilon \partial_\mu \P  \, , 
\\
\qeq c & =&  -\Sigma+ \Sigma_{\text{0}}\nn\,, \quad \quad \qquad 
\\ 
\qeq \bar{c} & =& b\,.\nn
\end{eqnarray}

\bibliographystyle{JHEP}
\bibliography{BibFile.bib}

\providecommand{\href}[2]{#2}\begingroup\raggedright\begin{thebibliography}{10}

\bibitem{Sen:2008vm}
A.~Sen, \emph{{Quantum Entropy Function from AdS(2)/CFT(1) Correspondence}},
  \href{https://doi.org/10.1142/S0217751X09045893}{\emph{Int. J. Mod. Phys. A}
  {\bfseries 24} (2009) 4225}
  [\href{https://arxiv.org/abs/0809.3304}{{\ttfamily 0809.3304}}].

\bibitem{correa:2019rdk}
D.H.~Correa, V.I.~Giraldo-Rivera and G.A.~Silva, \emph{{Supersymmetric mixed
  boundary conditions in AdS$_{2}$ and DCFT$_{1}$ marginal deformations}},
  \href{https://doi.org/10.1007/JHEP03(2020)010}{\emph{JHEP} {\bfseries 03}
  (2020) 010} [\href{https://arxiv.org/abs/1910.04225}{{\ttfamily
  1910.04225}}].

\bibitem{Gibbons:1976ue}
G.W.~Gibbons and S.W.~Hawking, \emph{{Action Integrals and Partition Functions
  in Quantum Gravity}},
  \href{https://doi.org/10.1103/PhysRevD.15.2752}{\emph{Phys. Rev. D}
  {\bfseries 15} (1977) 2752}.

\bibitem{Sen:2011ba}
A.~Sen, \emph{{Logarithmic Corrections to N=2 Black Hole Entropy: An Infrared
  Window into the Microstates}},
  \href{https://doi.org/10.1007/s10714-012-1336-5}{\emph{Gen. Rel. Grav.}
  {\bfseries 44} (2012) 1207}
  [\href{https://arxiv.org/abs/1108.3842}{{\ttfamily 1108.3842}}].

\bibitem{Bhattacharyya:2012wz}
S.~Bhattacharyya, B.~Panda and A.~Sen, \emph{{Heat Kernel Expansion and
  Extremal Kerr-Newmann Black Hole Entropy in Einstein-Maxwell Theory}},
  \href{https://doi.org/10.1007/JHEP08(2012)084}{\emph{JHEP} {\bfseries 08}
  (2012) 084} [\href{https://arxiv.org/abs/1204.4061}{{\ttfamily 1204.4061}}].

\bibitem{Sen:2012cj}
A.~Sen, \emph{{Logarithmic Corrections to Rotating Extremal Black Hole Entropy
  in Four and Five Dimensions}},
  \href{https://doi.org/10.1007/s10714-012-1373-0}{\emph{Gen. Rel. Grav.}
  {\bfseries 44} (2012) 1947}
  [\href{https://arxiv.org/abs/1109.3706}{{\ttfamily 1109.3706}}].

\bibitem{Sen:2012dw}
A.~Sen, \emph{{Logarithmic Corrections to Schwarzschild and Other Non-extremal
  Black Hole Entropy in Different Dimensions}},
  \href{https://doi.org/10.1007/JHEP04(2013)156}{\emph{JHEP} {\bfseries 04}
  (2013) 156} [\href{https://arxiv.org/abs/1205.0971}{{\ttfamily 1205.0971}}].

\bibitem{Keeler:2014bra}
C.~Keeler, F.~Larsen and P.~Lisbao, \emph{{Logarithmic Corrections to $N \geq
  2$ Black Hole Entropy}},
  \href{https://doi.org/10.1103/PhysRevD.90.043011}{\emph{Phys. Rev. D}
  {\bfseries 90} (2014) 043011}
  [\href{https://arxiv.org/abs/1404.1379}{{\ttfamily 1404.1379}}].

\bibitem{Karan:2020njm}
S.~Karan and B.~Panda, \emph{{Logarithmic corrections to black hole entropy in
  matter coupled $\mathcal{N} \geq 1$ Einstein-Maxwell supergravity}},
  \href{https://doi.org/10.1007/JHEP05(2021)104}{\emph{JHEP} {\bfseries 05}
  (2021) 104} [\href{https://arxiv.org/abs/2012.12227}{{\ttfamily
  2012.12227}}].

\bibitem{Banerjee:2021pdy}
G.~Banerjee and B.~Panda, \emph{{Logarithmic corrections to the entropy of
  non-extremal black holes in $ \mathcal{N} $ = 1 Einstein-Maxwell
  supergravity}}, \href{https://doi.org/10.1007/JHEP11(2021)214}{\emph{JHEP}
  {\bfseries 11} (2021) 214}
  [\href{https://arxiv.org/abs/2109.04407}{{\ttfamily 2109.04407}}].

\bibitem{Assel:2016pgi}
B.~Assel, D.~Martelli, S.~Murthy and D.~Yokoyama, \emph{{Localization of
  supersymmetric field theories on non-compact hyperbolic three-manifolds}},
  \href{https://doi.org/10.1007/JHEP03(2017)095}{\emph{JHEP} {\bfseries 03}
  (2017) 095} [\href{https://arxiv.org/abs/1609.08071}{{\ttfamily
  1609.08071}}].

\bibitem{GonzalezLezcano:2023cuh}
A.~Gonz\'alez~Lezcano, I.~Jeon and A.~Ray, \emph{{Supersymmetric localization:
  ${\cal N}=(2,2)$ theories on S$^2$ and AdS$_2$}},
  \href{https://arxiv.org/abs/2302.10370}{{\ttfamily 2302.10370}}.

\bibitem{Banerjee:2009af}
N.~Banerjee, S.~Banerjee, R.K.~Gupta, I.~Mandal and A.~Sen,
  \emph{{Supersymmetry, Localization and Quantum Entropy Function}},
  \href{https://doi.org/10.1007/JHEP02(2010)091}{\emph{JHEP} {\bfseries 02}
  (2010) 091} [\href{https://arxiv.org/abs/0905.2686}{{\ttfamily 0905.2686}}].

\bibitem{Dabholkar:2010uh}
A.~Dabholkar, J.~Gomes and S.~Murthy, \emph{{Quantum black holes, localization
  and the topological string}},
  \href{https://doi.org/10.1007/JHEP06(2011)019}{\emph{JHEP} {\bfseries 06}
  (2011) 019} [\href{https://arxiv.org/abs/1012.0265}{{\ttfamily 1012.0265}}].

\bibitem{Dabholkar:2011ec}
A.~Dabholkar, J.~Gomes and S.~Murthy, \emph{{Localization \& Exact
  Holography}}, \href{https://doi.org/10.1007/JHEP04(2013)062}{\emph{JHEP}
  {\bfseries 04} (2013) 062} [\href{https://arxiv.org/abs/1111.1161}{{\ttfamily
  1111.1161}}].

\bibitem{Gupta:2012cy}
R.K.~Gupta and S.~Murthy, \emph{{All solutions of the localization equations
  for N=2 quantum black hole entropy}},
  \href{https://doi.org/10.1007/JHEP02(2013)141}{\emph{JHEP} {\bfseries 02}
  (2013) 141} [\href{https://arxiv.org/abs/1208.6221}{{\ttfamily 1208.6221}}].

\bibitem{Dabholkar:2014wpa}
A.~Dabholkar, N.~Drukker and J.~Gomes, \emph{{Localization in supergravity and
  quantum $AdS_4/CFT_3$ holography}},
  \href{https://doi.org/10.1007/JHEP10(2014)090}{\emph{JHEP} {\bfseries 10}
  (2014) 090} [\href{https://arxiv.org/abs/1406.0505}{{\ttfamily 1406.0505}}].

\bibitem{LopesCardoso:2022hvc}
G.~Lopes~Cardoso, A.~Kidambi, S.~Nampuri, V.~Reys and M.~Rossell\'o, \emph{{The
  gravitational path integral for $ N=4$ BPS black holes from black hole
  microstate counting}},  \href{https://arxiv.org/abs/2211.06873}{{\ttfamily
  2211.06873}}.

\bibitem{Hristov:2021zai}
K.~Hristov and V.~Reys, \emph{{Factorization of log-corrections in
  AdS$_{4}$/CFT$_{3}$ from supergravity localization}},
  \href{https://doi.org/10.1007/JHEP12(2021)031}{\emph{JHEP} {\bfseries 12}
  (2021) 031} [\href{https://arxiv.org/abs/2107.12398}{{\ttfamily
  2107.12398}}].

\bibitem{deWit:2018dix}
B.~de~Wit, S.~Murthy and V.~Reys, \emph{{BRST quantization and equivariant
  cohomology: localization with asymptotic boundaries}},
  \href{https://doi.org/10.1007/JHEP09(2018)084}{\emph{JHEP} {\bfseries 09}
  (2018) 084} [\href{https://arxiv.org/abs/1806.03690}{{\ttfamily
  1806.03690}}].

\bibitem{Jeon:2018kec}
I.~Jeon and S.~Murthy, \emph{{Twisting and localization in supergravity:
  equivariant cohomology of BPS black holes}},
  \href{https://doi.org/10.1007/JHEP03(2019)140}{\emph{JHEP} {\bfseries 03}
  (2019) 140} [\href{https://arxiv.org/abs/1806.04479}{{\ttfamily
  1806.04479}}].

\bibitem{Iliesiu:2022kny}
L.V.~Iliesiu, S.~Murthy and G.J.~Turiaci, \emph{{Black hole microstate counting
  from the gravitational path integral}},
  \href{https://arxiv.org/abs/2209.13602}{{\ttfamily 2209.13602}}.

\bibitem{Gupta:2021roy}
R.K.~Gupta, S.~Murthy and M.~Sahni, \emph{{Quantum entropy of BMPV black holes
  and the topological M-theory conjecture}},
  \href{https://doi.org/10.1007/JHEP06(2022)053}{\emph{JHEP} {\bfseries 06}
  (2022) 053} [\href{https://arxiv.org/abs/2104.02634}{{\ttfamily
  2104.02634}}].

\bibitem{Ciceri:2023mjl}
A.~Ciceri, I.~Jeon and S.~Murthy, \emph{{Localization on $AdS_3 \times S^2$ I:
  the 4d/5d connection in off-shell Euclidean supergravity}},
  \href{https://arxiv.org/abs/2301.08084}{{\ttfamily 2301.08084}}.

\bibitem{David:2018pex}
J.R.~David, E.~Gava, R.K.~Gupta and K.~Narain, \emph{{Boundary conditions and
  localization on AdS. Part I}},
  \href{https://doi.org/10.1007/JHEP09(2018)063}{\emph{JHEP} {\bfseries 09}
  (2018) 063} [\href{https://arxiv.org/abs/1802.00427}{{\ttfamily
  1802.00427}}].

\bibitem{David:2019ocd}
J.R.~David, E.~Gava, R.K.~Gupta and K.~Narain, \emph{{Boundary conditions and
  localization on AdS. Part II. General analysis}},
  \href{https://doi.org/10.1007/JHEP02(2020)139}{\emph{JHEP} {\bfseries 02}
  (2020) 139} [\href{https://arxiv.org/abs/1906.02722}{{\ttfamily
  1906.02722}}].

\bibitem{Pittelli:2018rpl}
A.~Pittelli, \emph{{Supersymmetric localization of refined chiral multiplets on
  topologically twisted $H^2$ \texttimes{} $S^1$}},
  \href{https://doi.org/10.1016/j.physletb.2019.135154}{\emph{Phys. Lett. B}
  {\bfseries 801} (2020) 135154}
  [\href{https://arxiv.org/abs/1812.11151}{{\ttfamily 1812.11151}}].

\bibitem{Iannotti:2023jji}
D.~Iannotti and A.~Pittelli, \emph{{Twisted Index on Hyperbolic
  Four-Manifolds}},
  \href{https://doi.org/10.1007/s11005-024-01788-x}{\emph{Lett. Math. Phys.}
  {\bfseries 114} (2024) 39}
  [\href{https://arxiv.org/abs/2307.11634}{{\ttfamily 2307.11634}}].

\bibitem{Cabo-Bizet:2017jsl}
A.~Cabo-Bizet, V.I.~Giraldo-Rivera and L.A.~Pando~Zayas, \emph{{Microstate
  counting of AdS$_{4}$ hyperbolic black hole entropy via the topologically
  twisted index}}, \href{https://doi.org/10.1007/JHEP08(2017)023}{\emph{JHEP}
  {\bfseries 08} (2017) 023}
  [\href{https://arxiv.org/abs/1701.07893}{{\ttfamily 1701.07893}}].

\bibitem{Sen:2023dps}
A.~Sen, \emph{{Revisiting localization for BPS black hole entropy}},
  \href{https://arxiv.org/abs/2302.13490}{{\ttfamily 2302.13490}}.

\bibitem{Camporesi:1994ga}
R.~Camporesi and A.~Higuchi, \emph{{Spectral functions and zeta functions in
  hyperbolic spaces}}, \href{https://doi.org/10.1063/1.530850}{\emph{J. Math.
  Phys.} {\bfseries 35} (1994) 4217}.

\bibitem{Camporesi:1992tm}
R.~Camporesi, \emph{{The Spinor heat kernel in maximally symmetric spaces}},
  \href{https://doi.org/10.1007/BF02100862}{\emph{Commun. Math. Phys.}
  {\bfseries 148} (1992) 283}.

\bibitem{Camporesi:1995fb}
R.~Camporesi and A.~Higuchi, \emph{{On the Eigen functions of the Dirac
  operator on spheres and real hyperbolic spaces}},
  \href{https://doi.org/10.1016/0393-0440(95)00042-9}{\emph{J. Geom. Phys.}
  {\bfseries 20} (1996) 1}
  [\href{https://arxiv.org/abs/gr-qc/9505009}{{\ttfamily gr-qc/9505009}}].

\bibitem{GonzalezLezcano:2023uar}
A.~Gonz\'alez~Lezcano, I.~Jeon and A.~Ray, \emph{{Supersymmetry and
  complexified spectrum on Euclidean AdS$_2$}},
  \href{https://arxiv.org/abs/2305.12925}{{\ttfamily 2305.12925}}.

\bibitem{Osterwalder:1973zr}
K.~Osterwalder and R.~Schrader, \emph{{Euclidean fermi fields and a feynman-kac
  formula for boson-fermion models}}, {\emph{Helv. Phys. Acta} {\bfseries 46}
  (1973) 277}.

\bibitem{Osterwalder:1972vwp}
K.~Osterwalder and R.~Schrader, \emph{{Feynman-kac formula for euclidean fermi
  and bose fields}},
  \href{https://doi.org/10.1103/PhysRevLett.29.1423}{\emph{Phys. Rev. Lett.}
  {\bfseries 29} (1972) 1423}.

\bibitem{Gupta:2015gga}
R.K.~Gupta, Y.~Ito and I.~Jeon, \emph{{Supersymmetric Localization for BPS
  Black Hole Entropy: 1-loop Partition Function from Vector Multiplets}},
  \href{https://doi.org/10.1007/JHEP11(2015)197}{\emph{JHEP} {\bfseries 11}
  (2015) 197} [\href{https://arxiv.org/abs/1504.01700}{{\ttfamily
  1504.01700}}].

\bibitem{Murthy:2013xpa}
S.~Murthy and V.~Reys, \emph{{Quantum black hole entropy and the holomorphic
  prepotential of N=2 supergravity}},
  \href{https://doi.org/10.1007/JHEP10(2013)099}{\emph{JHEP} {\bfseries 10}
  (2013) 099} [\href{https://arxiv.org/abs/1306.3796}{{\ttfamily 1306.3796}}].

\bibitem{Bianchi:2001kw}
M.~Bianchi, D.Z.~Freedman and K.~Skenderis, \emph{{Holographic
  renormalization}},
  \href{https://doi.org/10.1016/S0550-3213(02)00179-7}{\emph{Nucl. Phys. B}
  {\bfseries 631} (2002) 159}
  [\href{https://arxiv.org/abs/hep-th/0112119}{{\ttfamily hep-th/0112119}}].

\bibitem{Gradshteyn:1702455}
I.S.~Gradshteyn, I.M.~Ryzhik, D.~Zwillinger and V.~Moll, \emph{{Table of
  integrals, series, and products; 8th ed.}}, Academic Press, Amsterdam (Sep,
  2014), \href{https://doi.org/0123849330}{0123849330}.

\end{thebibliography}\endgroup
\end{document}